\documentclass[12pt]{article}
\pdfoutput=1

\usepackage{putex}
\usepackage{graphicx}
\usepackage{caption}
\usepackage{amsmath}
\usepackage{array}
\usepackage{subcaption}
\usepackage{epstopdf}
\usepackage{enumerate}
\usepackage{cite}
\usepackage{youngtab}
\usepackage{tensor}
\usepackage{slashed}
\usepackage{multirow}
\usepackage[aligntableaux=center]{ytableau}
\usepackage{rotating}
\usepackage[utf8]{inputenc}
\usepackage{comment}
\usepackage[
      colorlinks=true,
      linkcolor=blue,
      urlcolor=blue,
      filecolor=black,
      citecolor=red,
      ]{hyperref}

\def\W{{\cal W}}\def\c{\cite}

\def\t{\tau}
\def\flim{\lim_{s,t\rar\i}}
\def\z{\zeta}

\def\eps{\epsilon}
\def\Nc{{\cal N}}
\def\Lads{L_{\rm AdS}}
\def\Ac{{\cal A}}
\def\b{\beta}
\def\M{{\cal M}}
\def\zb{\overline{z}}
\def\rar{\rightarrow}
\def\F{{\cal F}}

\def\winf{{\cal W}_{\infty}}

\def\lp{\ell_{11}}
\def\Mt{\widetilde{M}}

\def\flat{{\rm flat}}
\def\tree{\rm tree}

\def\l{\left}

\def\o{\over}
\def\Lc{\mathcal{L}}

\def\eqr{\eqref}
\def\sec{\section}
\def\Oc{{\cal O}}
\def\l{\lambda}
\def\vs{\vskip .1 in}
\def\a{\alpha}
\def\O{{\cal O}}

\def\D{\Delta}
\def\del{\delta}
\def\g{\gamma}
\def\s{\sigma}
\def\ssec{\subsection}
\def\sssec{\subsubsection}
\def\i{\infty}
\def\foot{\footnote}
\def\Dc{{\cal D}}
\def\Gc{{\cal G}}

\newcommand {\be} {\begin {equation}}
\newcommand {\ee} {\end {equation}}

\newcommand {\bes} {\begin {equation*}}
\newcommand {\ees} {\end {equation*}}

\newcommand{\e}[2] {\begin{equation} \label{#1} #2 \end{equation}}
\newcommand{\es}[2] {\begin{equation} \label{#1} \begin{split} #2 \end{split} \end{equation}}

\newcommand{\R}{\mathbb{R}}

\newcommand{\cO}{{\mathcal O}}

\newcommand{\beq}{\begin{equation}}
\newcommand{\eeq}{\end{equation}}

\def\p{\partial}

\def\ie{\begin{equation}\begin{aligned}}
\def\fe{\end{aligned}\end{equation}}

\newcommand{\la}{\langle}
\newcommand{\ra}{\rangle}

\newcommand{\m}{\mu}

\newcommand{\A}{{\alpha}}
\newcommand{\B}{{\beta}}

\numberwithin{equation}{section}

\newcommand\Tstrut{\rule{0pt}{2.3ex}}       
\newcommand\Bstrut{\rule[-1.3ex]{0pt}{0pt}} 
\newcommand\TBstrut{\Tstrut\Bstrut}         

\def\<{\langle}
\def\>{\rangle}

\begin{document}

\preprint{PUPT-2562}

\institution{PU}{Joseph Henry Laboratories, Princeton University, Princeton, NJ 08544, USA}
\institution{CU}{Walter Burke Institute for Theoretical Physics, Caltech, Pasadena, CA 91125, USA }

\title{M-Theory Reconstruction from (2,0) CFT \\and the Chiral Algebra Conjecture}

\authors{Shai M.~Chester\worksat{\PU} and Eric Perlmutter\worksat{\CU}}

\abstract{We study various aspects of the M-theory uplift of the $A_{N-1}$ series of $(2,0)$ CFTs in 6d, which describe the worldvolume theory of $N$ M5 branes in flat space. We show how knowledge of OPE coefficients and scaling dimensions for this CFT can be directly translated into features of the momentum expansion of M-theory. In particular, we develop the expansion of the four-graviton S-matrix in M-theory via the flat space limit of four-point Mellin amplitudes. This includes correctly reproducing the known contribution of the $R^4$ term from 6d CFT data. Central to the calculation are the OPE coefficients for half-BPS operators not in the stress tensor multiplet, which we obtain for finite $N$ via the previously conjectured relation \cite{Beem:2014kka} between the quantum ${\cal W}_N$ algebra and the $A_{N-1}$ $(2,0)$ CFT. We further explain how the $1/N$ expansion of ${\cal W}_N$ structure constants exhibits the structure of protected vertices in the M-theory action. Conversely, our results provide strong evidence for the chiral algebra conjecture.}
\date{}

\maketitle

\setcounter{tocdepth}{2}
\tableofcontents

\section{Introduction}
\label{intro}

The goal of this paper is to develop a quantitative study of M-theory by way of its holographic duality to six-dimensional conformal field theory (CFT) with maximal (2,0) supersymmetry, using modern results from the conformal bootstrap and techniques for computing correlation functions in large $N$ CFTs. 

The $A_{N-1}$ (2,0) CFT, on which we will focus, has various descriptions (e.g. \cite{Witten:1995zh,Seiberg:1997ax, Aharony:1997th,Lambert:2010iw,Douglas:2010iu}). Perhaps the most profitable is its realization as the worldvolume theory of $N$ M5 branes in flat space, whose gravitational backreaction generates an AdS$_7\times S^4$ solution of M-theory. AdS/CFT then provides the usual dictionary for computing various observables in the $1/N$ expansion \c{Maldacena:1997re,Gubser:1998bc,Witten:1998qj}; indeed, the notion of a well-defined $1/N$ expansion was first made explicit by the existence of the bulk dual. Despite being a non-Lagrangian, non-gauge theory with somewhat mysterious origins and $O(N^3)$ degrees of freedom at large $N$, the (2,0) CFT behaves similarly in many respects to lower-dimensional gauge theories that furnish canonical examples of AdS/CFT, such as the 4d $\Nc=4$ super-Yang-Mills (SYM) duality to AdS$_5\times S^5$. 

On the other hand, given our utter lack of a complete description of M-theory, the bulk is not terribly useful for determining finite $N$ aspects of the dual CFT. However, we can turn this problem around using the modern perspective of the conformal bootstrap, which gives an {\it a priori} independent formulation of the (local sector of the) CFT. This provides an independent tool for constructing M-theory at the non-perturbative level, a philosophy that we will substantiate in this work. 

An initial implementation of the numerical bootstrap to the (2,0) CFT was performed in \cite{Beem:2015aoa}, which led to the first predictions for finite $N$ data for low-lying non-BPS operators that appear in the stress tensor operator product expansion (OPE). More relevant for us will be the remarkable analytic progress in the BPS sector. The half-BPS supermultiplets in interacting theories have bottom components $S_k$ with $k=2,3,\ldots$\footnote{The $k=1$ case only exists for the free theory.} and conformal dimension $\D_k=2k$, which are traceless symmetric tensors of the $\mathfrak{so}(5)_R$ symmetry. The KK reduction on AdS$_7\times S^4$ \c{Pilch:1984xy,vanNieuwenhuizen:1984iz,Nastase:1999cb,Nastase:1999kf} identifies the $S_k$ (modulo mixing) with scalar fields $\phi_k$ in AdS, of squared mass $(mL_{\rm AdS})^2 = 2k(2k-6)$, which uplift to admixtures of the 11d graviton and three-form potential with legs on $S^4$. While $\D_k$ is independent of $N$, the OPE coefficients $\l_{k_1k_2k_3}$ are not. In \cite{Beem:2014kka}, it was conjectured that these OPE coefficients sit in one-to-one correspondence with the structure constants $C_{k_1k_2k_3}$ of the well-studied two-dimensional $\W_N$ chiral algebra, with the auspicious central charge assignment $c=4N^3-3N-1$. This algebra is freely generated by an infinite tower of conserved currents $W_k$ of spins $s=2,3,\ldots,N$, which lie in correspondence with the half-BPS operators $S_k$ mentioned above.

The $\W_N$ chiral algebra conjecture is powerful: it determines, in principle, an infinite number of OPE coefficients of the (2,0) CFT. Many of the $\W_N$ structure constants, which are completely determined by the Jacobi identities, are also explicitly known.\foot{As we review below, there is a choice of basis in which there is a conjecture for all of them \c{Prochazka:2014gqa}; in the most physical basis for 6d purposes, the first many low-lying ones are known \c{Gaberdiel:2012ku}.} This data is highly quantum from the M-theory perspective, as it is known in closed form for finite $c$ and finite $N$, unlike the currently known analogous results for protected operator algebras in $d=3,4$ maximally-supersymmetric CFTs \c{Beem:2013sza,Chester:2014mea} (some of which, however, do admit finite-dimensional integral representations \c{Dedushenko:2016jxl}). In \cite{Beem:2014kka}, it was shown that the $\W_N$ OPE coefficients with $c\approx 4N^3$ in the large $N$ limit correctly reproduce previous computations of tree-level three-point functions in the (2,0) CFT as computed from AdS$_7\times S^4$ \c{Bastianelli:1999en,Campoleoni:2011hg}. Some further aspects of the conjecture were substantiated in \c{Cordova:2016cmu} using localization and the $\W_N$ chiral algebra of $A_{N-1}$ Toda CFT. One aim of this paper is to test this chiral algebra conjecture beyond leading order in $1/N$; as we explain below, we find strong evidence, both perturbative and non-perturbative, that the conjecture is indeed correct. 

Before explaining what exactly we will compute, let us set the target. Even putting aside the deeper non-perturbative aspects of M-theory, the expansion of the 11d four-point superamplitude, $\Ac^{11}$, is not well understood. The 11d amplitude takes the form \c{Grisaru:1976vm}
 \e{A11d}{\Ac^{11}(p_i;\z_i) = f(s,t) {\cal A}^{11}_{ R, \tree}(p_i;\z_i) \,.} 
${\cal A}^{11}_{R, \tree}$ is the supergravity tree-level amplitude, 
\e{atree}{{\cal A}^{11}_{ R, \tree}(p_i;\z_i) =  \lp^9\widehat K {2^6\o stu}}
where $\widehat K$ is an overall universal kinematic factor (whose form we later recall) that is a function of graviton polarization vectors $\zeta^{\mu}$ and momenta $p^{\mu}$, and $(s,t,u)$ are the 11d Mandelstam variables. The function $f(s,t)$, dependent on the Mandelstam variables only, encodes the momentum expansion,
  \es{fexp2}{ f(s, t) &= 1 + \ell_{11}^6 f_{R^4}(s, t)+ \ell_{11}^9 f_{\rm 1-loop}(s,t)  +  \ell_{11}^{12} f_{D^6 R^4}(s, t)     +  \ell_{11}^{14} f_{D^8 R^4}(s, t) \\ &   + \ell_{11}^{15} f_{{\rm 1-loop},R^4}(s,t)+  \ell_{11}^{16} f_{D^{10} R^4}(s, t)+ \ell_{11}^{18} f_{\rm 2-loop}(s, t)  + \ell_{11}^{18} f_{D^{12}R^4}(s,t)  + \cdots    \,,  } 
where all 11d loop corrections come in powers of $\ell_{11}^9$ times the tree-level vertices.  Among the non-loop terms, only the $R^4$ and $D^6R^4$ terms are known from previous computations, as reviewed in Appendix \ref{appstring}.\foot{The precise tensor appearing at $R^4$ is $t_8t_8R^4$, plus $\eps_{11}$ terms that do not contribute to the four-graviton amplitude. Further details about $R^4$ and its superpartners may be found in e.g. \c{Tseytlin:2000sf}.} At loop-level, 1- and 2-loop amplitudes
are known from 11d supergravity computations \c{Green:1997as,Green:1999pu}. Beginning at $D^8R^4$, the vertices are no longer protected by supersymmetry and their coefficients are not known, although there exist conjectures in the literature \c{Russo:1997mk}. It is of great interest to improve on this state of affairs -- specifically, the outstanding problem of determining $D^8R^4$ and beyond, and of unveiling the finite $N$ spectrum of M-theory -- by computing the CFT four-point functions $\la S_kS_kS_kS_k\ra$, and uplifting them to M-theory. This would be a remarkable holographic window onto the perturbative structure of M-theory and, by compactification, type IIA string theory. 

In this paper, we will articulate a concrete strategy for doing this. As a step toward the longer-term goal of $D^8R^4$, we will explicitly demonstrate this strategy by deriving the $R^4$ term in \eqr{fexp2} from CFT, as recently done in a closely related context using AdS$_4\times S^7$ and the ABJM CFT \cite{Chester:2018aa}. Let us summarize the idea. We first compute $\la S_kS_kS_kS_k\ra$ in Mellin space in the $1/c$ expansion by solving the 6d superconformal Ward identity and using independent CFT data to fix any free parameters in the solution. This fixing relies crucially on input from $\W_N$ to fix half-BPS structure constants. We then use the flat space limit formula for Mellin amplitudes \cite{Penedones:2010ue} to relate this correlator at a given order in $1/c$ to terms in the $\ell_{11}\ll 1$ expansion of $\Ac^{11}$, using the holographic relation
 \e{LellpRelation}{  \left(\frac{\Lads}{\ell_{11}}\right)^9 \approx 16c+O(c^0) \,. }
where $L_{S^4} = L_{{\rm AdS}}/2$.\foot{In defining $\ell_{11}$, we use the conventions of \c{Tseytlin:2000sf}. The relation between $L_{\rm AdS}$ and $L_{S^4}$ is believed to hold to all orders in $\ell_{11}$ \c{Kallosh:1998qs}.} The direct relation of the $1/c$ expansion to the $\ell_{11}$ expansion follows from dimensional analysis in the reduction on AdS$_7\times S^4$, and the absence of a dimensionless coupling in M-theory. One novelty of the 6d case is that (for reasons explained below) in order to fix the parameters necessary to reproduce the $R^4$ coefficient from presently known (2,0) CFT data, we will need to study the $k=3$ correlator, as opposed to the stress tensor multiplet correlator ($k=2$). Along the way, we will explain how to uplift $\la S_kS_kS_kS_k\ra$ to 11d for arbitrary $k$. 
\vs

In {\bf Section \ref{4pointsec}}, we review the basic features of the (2,0) CFT and the implications of superconformal symmetry on the structure of four-point functions of the half-BPS superconformal primaries $S_k$. We then recall the conjectured relation between the $\W_N$ algebra and $A_{N-1}$ (2,0) CFT data. Using properties of $\W_N$, we show that all half-BPS OPE coefficients $\l_{k_1k_2k_3}$ admit a $1/c$ expansion of the form
\e{lm123}{\l_{k_1k_2k_3}^2 = c^{-1}F_R(c) + c^{-5/3} F_{R^4}(c) + c^{-7/3}F_{D^6R^4}(c)\,,}
where the $F_i(c)$ have an expansion in non-negative integer powers of $1/c$. This structure is precisely what one expects from M-theory: in particular, it is consistent with the only protected 11d vertices being $R$, $R^4$ and $D^6R^4$ (hence the subscripts), with the $F_i(c)$ encoding bulk loops in the presence of these terms. The fact that $\l_{k_1k_2k_3}^2$ obeys the form \eqr{lm123}, for any $k_i$, is strong all-orders evidence for the validity of the identification of $\W_N$ with central charge $c=4N^3-3N-1$ as the chiral algebra of the (2,0) CFT. Conversely, this may be viewed as suggestive evidence of the absence of 10- and 12-derivative terms in 11d ($\sim D^2R^4$ and $D^4R^4$ + superpartners).\foot{In the dimensional reduction on AdS$_7\times S^4$, cancellations are possible among different putative terms at a fixed derivative order, e.g. $D^4R^4$ and $R^6$. However, since \eqr{lm123} describes the behavior of {\it all} half-BPS three-point functions, consistency of a nonzero 10- or 12-derivative action with \eqr{lm123} would require an infinite number of cancellations.} 

In {\bf Section \ref{hol}}, we study the four-point functions $\la S_kS_kS_kS_k\ra$. We work with the corresponding Mellin amplitudes, which we denote $M_k$. After writing their general form, we explore the space of solutions to the 6d superconformal Ward identity, focusing especially (but not exclusively) on $k=2,3$. The solutions are distinguished by whether they are meromorphic or polynomial, and are organized according to their degree in the limit of $s,t\rar\i$. 

In {\bf Section \ref{s5}}, we give a physical analysis of the solution space in Section 3 and explain how to uplift to M-theory. We first show how to extract the 11d flat-space amplitude $\Ac^{11}$ from the $1/c$ expansion of 6d Mellin amplitudes $M_k$ at large $s,t$, for any $k$. This involves an adaptation of Penedones' formula to the case of arbitrary KK modes. (See \eqr{FlatLimit}, \eqr{11damp}.) A nice feature of this procedure is that, in addition to producing the function $f(s,t)$, the overall kinematic factor $\widehat K$ of $\Ac^{11}$ can be seen to follow quite directly from the flat space limit of the 6d superconformal Ward identity itself. (See \eqr{scwardflat}.) Moreover, the same factor $\Theta_4^{\flat}(s,t;\s,\tau)$ appears in the 4d, $\Nc=4$ superconformal Ward identity. Therefore, the flat space limit of 4d $\Nc=4$ SYM four-point functions implies that type IIB string amplitudes are proportional to the universal $\widehat K$ factor to all loop orders. This has sometimes been indirectly argued on general grounds (e.g. \c{Green:2008bf}), and $\widehat K$ is known to appear in type II string theory through three-loop order \c{DHoker:2005vch,Gomez:2013sla}; here we give a rigorous derivation of its appearance to all orders in type IIB.

With this understanding, we explain how the coefficients of the Mellin amplitudes are directly related to CFT data, namely, OPE coefficients and scaling dimensions. The main physical point is that the degree of the solutions is correlated with the order in $1/c$ at which they first appear in CFT; in particular, a degree-$p$ solution $M_k^{(p)}$ has scaling $c^{-(2p+7)/9}$ to leading order in $1/c$. This is shown to follow from the flat-space limit and the absence of a dimensionless coupling in M-theory. This allows us to explain, physically, some features of the Mellin amplitudes found in Section \ref{hol}. The result may be viewed as an M-theory version of previous arguments relating coefficients of solutions of crossing to powers of the higher spin gap in large $N$ CFTs \c{Alday:2014tsa,Caron-Huot:2017vep}. A related perspective on this $c$-scaling is given in terms of the dimensional reduction of M-theory on AdS$_7\times S^4$. Together with previous knowledge of the M-theory amplitude through 14-derivative order, we can rule out candidate polynomial solutions of crossing symmetry at $O(c^{-17/9})$ and $O(c^{-19/9})$. (A similar argument was made in \c{Chester:2018aa}.) This last statement, which uses  general features of KK reduction on AdS$\times \M$, applies to any CFT with an M-theory dual of this form. We also present a sharp signature of the four-point functions of putative large $c$ CFTs with a hierarchy between the AdS and KK scales, $\Lads \gg L_\M$.

In {\bf Section \ref{compare}}, we put everything together to develop the precise dictionary between M-theory and (2,0) CFT. First, we derive the $R^4$ coefficient via the $k=3$ four-point function at $O(c^{-5/3})$.  This is possible because the $k=3$ amplitude at $O(c^{-5/3})$ happens to be determined by one free parameter, which we can take to be the OPE coefficient $\lambda_{ 334}$. This is in turn fixed by $\W_N$. The result perfectly matches the M-theoretic prediction,
\e{fr4}{f_{R^4}(s,t) = \frac{stu}{3 \cdot 2^7}~.}
We note that the $O(c^{-5/3})$ term in $\lambda_{ 334}$ is extracted from a $c^{-1}N^{-2}$ term in $\W_N$; in particular, one does not need to know the sub-leading $O(N)$ term in $c$, which descends from the 11d $R^4$ term in the first place \cite{Tseytlin:2000sf}. Turning next to higher order terms $\sim D^{2m}R^4$, $\W_N$ does not provide enough constraints on the $k=3$ amplitude to completely fix the solutions. (This is due to the existence of pure polynomial solutions.) Instead, our strategy will be to relate the higher degree Mellin amplitude coefficients to CFT data that is not determined by $\W_N$ -- namely, anomalous dimensions of unprotected double-trace operators and OPE coefficients of protected operators that do not live in $\W_N$. Thus, future constraints on this data can be translated into constraints on M-theory amplitudes. Because the number of Mellin amplitudes grows with $k$, we will focus on the lowest case $k=2$. The output of this procedure is given in Table \ref{resultList}. 

In {\bf Section \ref{conc}}, we conclude with some future directions.

Several Appendices complement the main text. These include technical details on Mellin amplitudes, superconformal Ward identities, superconformal blocks, the OPE of two stress tensor multiplet scalars $S_2$, and a review of the derivation of $f_{R^4}(s,t)$ and $f_{D^6R^4}(s,t)$ from the uplift of type IIA string theory.

\sec{(2,0) Correlators and the $\W_N$ Chiral Algebra}
\label{4pointsec}
Let us begin by briefly reviewing the spectrum of operators in the $A_{N-1}$ series (2,0) CFT, with superconformal algebra $\mathfrak{osp}(8^*|4)\supset  \mathfrak{so}(2)\oplus\mathfrak{so}(6)\oplus \mathfrak{so}(5)_R$. For operators in the traceless symmetric spin-$j$ irrep of the $\mathfrak{so}(6)$ Lorentz algebra, we denote their quantum numbers under the bosonic subalgebra as
\e{}{(\Delta,j)_{[k1\,k2]\,,}
}
where $[k_,k_2]$ are $\mathfrak{so}(5)_R$ Dynkin labels. The CFT contains a single tower of half-BPS superconformal primaries, living in the $(2k,0)_{[k\,0]}$, 
with  $k=2,3,\ldots$. We denote their superconformal multiplet as $\Dc{[k0]}$. We can view these operators as the rank-$k$ symmetric traceless products of the ${\bf5}$, so we can denote them as traceless symmetric tensors $S_{I_1\dots I_k}( x)$ of $\mathfrak{so}(5)$, where $I_i=1,\dots5$. It is convenient and conventional to contract with an auxiliary polarization vector $Y^I$ that is constrained to be null, $Y_i\cdot Y_i=0$, so that
\e{S}{
S_k( x,Y)\equiv S_{I_1\dots I_k}Y^{I_1}\cdots Y^{I_k}\,.
}

We will often perform explicit computations involving the two lowest multiplets. The $k=2$ multiplet, $\Dc[20]$, is the stress tensor multiplet, whose bottom component is a scalar with $\Delta=4$ in the ${\bf 14}$ of $\mathfrak{so}(5)$, and appears in all local $(2,0)$ SCFTs. The next lowest half-BPS multiplet, $\Dc[30]$, has a scalar bottom component with $\Delta=6$ in the ${\bf30}$ of $\mathfrak{so}(5)$. The stress tensor itself $T_{\mu\nu}$ has a two-point function
\e{}{ \la T_{\mu\nu}(x)T_{\rho\sigma}(0) \ra = c_T {{\cal I_{\mu\nu\rho\sigma}}(x)\o | x|^{12}}\,,}
where ${\cal I}_{\mu\nu\rho\sigma}(x)$ is a fixed tensor structure whose form can be found in \c{Osborn:1993cr}. The coefficient $c_T$ is proportional to the unique $c$-type central charge appearing in the (2,0) conformal anomaly \c{Bastianelli:2000hi}. A free (2,0) tensor multiplet may be taken to have $c=1$, while a $(2,0)$ theory labeled by Lie algebra $\mathfrak{g}$ has central charge $c(\mathfrak{g})=4d_\mathfrak{g}h^\vee_\mathfrak{g}+r_\mathfrak{g}$, where $d_\mathfrak{g}$, $h^\vee_\mathfrak{g}$, and $r_\mathfrak{g}$ are the dimension, dual Coxeter number, and rank of $\mathfrak{g}$, respectively. For the $A_{N-1}$ series of interest here,
\e{cT66}{c(A_{N-1})=4N^3-3N-1\,.}
These results for $c$ were first motivated by R-symmetry anomaly and holographic computations \cite{Harvey:1998bx, Tseytlin:2000sf, Intriligator:2000eq}, conjectured in \cite{Beem:2015aoa}, and proven in \c{Cordova:2016cmu}. (See also \cite{Beccaria:2015ypa,Cordova:2015vwa,Cordova:2015fha,Beccaria:2017dmw,Yankielowicz:2017xkf,Chang:2017xmr} for some recent related results about the $c$-type anomaly in 6d SCFT.)

\ssec{Half-BPS Four-Point Functions}

Conformal symmetry and $\mathfrak{so}(5)$ symmetry implies that the four point function of $S_k( x,Y)$ takes the form
\es{4point}{
\langle S_k( x_1,Y_1)S_k( x_2,Y_2)S_k( x_3,Y_3)S_k( x_4,Y_4) \rangle=\frac{(Y_1\cdot Y_2)^k(Y_3\cdot Y_4)^k}{| x_{12}|^{4k}| x_{34}|^{4k}}\mathcal{G}_k(U,V;\sigma,\tau)\,,
}
where $U$ and $V$ are conformally-invariant cross ratios and $\sigma$ and $\tau$ are $\mathfrak{so}(5)$ invariants formed out of the polarizations:
 \es{uvsigmatauDefs}{
  U \equiv \frac{{x}_{12}^2 {x}_{34}^2}{{x}_{13}^2 {x}_{24}^2} \,, \qquad
   V \equiv \frac{{x}_{14}^2 {x}_{23}^2}{{x}_{13}^2 {x}_{24}^2}  \,, \qquad
   \sigma\equiv\frac{(Y_1\cdot Y_3)(Y_2\cdot Y_4)}{(Y_1\cdot Y_2)(Y_3\cdot Y_4)}\,,\qquad \tau\equiv\frac{(Y_1\cdot Y_4)(Y_2\cdot Y_3)}{(Y_1\cdot Y_2)(Y_3\cdot Y_4)} \,,
 }
 with $x_{ij}\equiv x_i-x_j$. Since \eqref{4point} is a degree $k$ polynomial in each $Y_i$ separately, the quantity $\mathcal{G}_k(U,V;\sigma,\tau)$ is a degree $k$ polynomial in $\sigma$ and $\tau$.

So far, we have imposed the bosonic subgroups of the $\mathfrak{osp}(8^*|4)$ algebra. The constraints from the fermionic subgroups are captured by the superconformal Ward identities \cite{Dolan:2004mu}, which can be expressed as differential operators on all four arguments of $\mathcal{G}_k(U,V;\sigma,\tau)$ whose explicit form we review in Appendix \ref{SUSYWARD}. For 6d $(2,0)$ SCFTs, there are two ways of satisfying these constraints. 

In the first method, which can actually be used in any dimension, we decompose $ \mathcal{G}_k(U,V;\sigma,\tau)$ into superconformal blocks $\mathfrak{G}_{\mathcal{M}}$ by taking the OPE twice in \eqref{4point}, which yields 
\es{SBDecomp}{
     \mathcal{G}_k(U,V;\sigma,\tau)=\sum_{\mathcal{M}_k\in\mathfrak{osp}(8^*|4)}\lambda^2_{k,\mathcal{M}_k}\mathfrak{G}_{\mathcal{M}_k}(U,V;\sigma,\tau)\,,
}
where ${\cal M}_k$ runs over all $\mathfrak{osp}(8^*|4)$ multiplets appearing in the $S_k \times S_k$ OPE, and $\lambda^2_{k,{\mathcal{M}}_k}$\footnote{When $\mathcal{M}_k=\mathcal{D}[k'0]$, these squared OPE coefficients were denoted in the introduction as $\lambda^2_{kkk'}$.} is the OPE coefficient squared for each such supermultiplet $\mathcal{M}_k$. The selection rules for the OPE of half-BPS multiplets have been worked out in \cite{Heslop:2004du,Ferrara:2001uj} and were summarized for general $k$ in \cite{Beem:2015aoa}. The supermultiplets that appear in a four point function of identical $S_k$'s are
\es{opemultEq}{
&S_k\times S_k=\sum_{m=0}^k\sum_{n=0}^{k-m}\mathcal{D}[2(k-m-n),2n]\\
&+\sum_{m=1}^k\left[\sum_{n=0,2,\dots}^{k-m}\,\sum_{j=0,2,\dots}^\infty\mathcal{B}[2(k-m-n),2n]_j+\sum_{n=1,3,\dots}^{k-m}\,\sum_{j=1,3,\dots}^\infty\mathcal{B}[2(k-m-n),2n]_j\right]\\
&+\sum_{m=2}^k\left[\sum_{n=0,2,\dots}^{k-m}\sum_{j=0,2,\dots}^\infty\mathcal{A}[2(k-m-n),2n]_{\Delta,j}+\sum_{n=1,3,\dots}^{k-m}\sum_{j=1,3,\dots}^\infty\mathcal{A}[2(k-m-n),2n]_{\Delta,j}\right],\\
}
where the spins $j$ refer to rank-$j$ traceless symmetric irreps of $\mathfrak{so}(6)$ with Dynkin labels $[j00]$, which are the only irreps that can appear, and for interacting SCFTs we should further remove the $\mathcal{B}[00]_j$ multiplet, which contains conserved currents that only appear in the free theory. The scaling dimensions of bottom components of the supermultiplets in \eqref{opemultEq} are
\es{dims}{
\mathcal{D}[k_1k_2]&: \quad\Delta=2(k_1+k_2)\,,\\
\mathcal{B}[k_1k_2]_j&:\quad\Delta=2(k_1+k_2)+4+j\,,\\
\mathcal{A}[k_1k_2]_{\D,j}&:\quad \Delta\geq 2(k_1+k_2)+6+j\,.\\
}
The $\mathcal{A}$ multiplets that appear here are unprotected, while the rest are annihilated by some fraction of supercharges, and so have fixed dimension. The $\mathcal{D}[k0]$ are the half-BPS multiplets whose bottom component we called $S_k$, and $k$ of these multiplets appear in $S_k\times S_k$. The lowest such multiplet is always the stress tensor multiplet $\mathcal{D}[20]$, whose OPE coefficient is 
\es{cTolam}{
\lambda^2_{k,\mathcal{D}[20]}=\frac{k^2}{c}\,.
}
This follows from Wick contractions and the normalization $c=1$ for a free tensor multiplet.

Each superconformal block $\mathfrak{G}_{{\mathcal{M}}_k}$ in \eqref{SBDecomp} receives contributions from conformal primaries with different spins $j'$, scaling dimensions $\Delta'$, and $\mathfrak{so}(5)$ irreps $[2b\, 2(a-b)]$ for $a=0,1,\dots k$ and $b=0,\dots, a$ that appear in the tensor product $[k0]\otimes [k0]$. The superconformal block can thus be written as a linear combination of the conformal blocks $G_{\Delta',j'}$\footnote{We normalize our conformal blocks as $
\lim_{U\to0,V\to1}G_{\Delta',j'}=U^\frac{\Delta'-j'}{2}(1-V)^{j'}$.} corresponding to the conformal primaries in $\mathcal{M}_k$ as
 \es{GExpansion}{
  {\mathfrak G}_{\mathcal{M}_k}(U, V; \sigma, \tau) = \sum_{a=0}^k \sum_{b = 0}^a Y_{ab}(\sigma, \tau)  \sum_{(\Delta',j')\in\mathcal{M}_k} A^{\mathcal{M}_k}_{ab \Delta' j'}(\Delta, j) G_{\Delta',j'}(U,V) \,,
 }  
where the polynomials $Y_{ab}(\sigma, \tau)$ are eigenfunctions of the $\mathfrak{so}(5)$ Casimir for irrep $[2b\, 2(a-b)]$ of maximal degree $k$. For general $k$, these can be computed using Appendix D in \cite{Nirschl:2004pa}, and we list the explicit forms for $k=2,3$ in Appendix \ref{SUSYWARD}. The $A^{\mathcal{M}_k}_{ab\Delta'j'}(\Delta, j)$ are rational functions of $\Delta$ and $j$. For $k=2$, we work out some of these coefficients in Appendix \ref{super20}.\footnote{These superconformal blocks have also been derived in different bases in \cite{Dolan:2004mu,Heslop:2004du,Beem:2015aoa}.}

The second way of imposing the constraints from the superconformal Ward identity is special to 6d $(2,0)$ SCFTs. We can satisfy the Ward identities by writing $\mathcal{G}_k$ as
\es{Ward2}{
\mathcal{G}_k(U,V;\sigma,\tau)=\mathcal{F}_k(U,V;\sigma,\tau)+\Upsilon\circ \mathcal{H}_k(U,V;\sigma,\tau)\,,
}
where $\Upsilon$ is a complicated differential operator whose explicit form is given in Appendix \ref{SUSYWARD}, and $\mathcal{F}_k(U,V;\sigma,\tau)$ and $\mathcal{H}_k(U,V;\sigma,\tau)$ are degree $k$ and $k-2$ in $\sigma,\tau$, respectively. These functions are defined so that only $\mathcal{F}_k(U,V;\sigma,\tau)$ contributes to the 2d chiral algebra four-point function, which we will describe in the next subsection.

\subsection{The $\W_N$ Chiral Algebra and AdS$_7\times S^4$}
\label{chiral}

It was conjectured in \cite{Beem:2014kka} that in every (2,0) CFT labeled by $\mathfrak{g}$, the OPE data of half-BPS operators and a subset of their protected composites -- specifically, among those appearing in $S_k\times S_k$, the $\Dc[2n\,0]$ and ${\cal B}[2n\,0]$ multiplets -- are determined by the dynamics of a 2d chiral algebra, ${\cal W}_{\mathfrak{g}}$. In the case $\mathfrak{g}=A_{N-1}$, the algebra is the well-known $\W_N$ algebra. This algebra is generated by a finite tower of holomorphic currents, $W_k(z)$, of integer spins $k=2,3,\ldots,N$, and depends on a single free parameter $c_{2d}$, the central charge. The conjecture stipulates that
\e{cN}{c_{2d}=c(A_{N-1})=4N^3-3N-1~.}
$\W_N$ may be obtained as the quotient of the infinitely-generated $\winf[N]$ algebra: $\W_N\simeq  \winf[N]/\chi_N$, where $\chi_N$ is the ideal formed by all generators of spins $s>N$, and 
$\winf[N]$ is the so-called ``quantum algebra'', to be distinguished from the ``classical algebra'' which is defined as the $c\rar\i$ limit of $\winf[\mu]$ with fixed $\mu$ \c{Pope:1989ew,Pope:1989sr}. Henceforth we will refer to the central charge simply as $c$, and write the $\W_N$ OPE as
\e{}{W_i(z)W_j(0) \sim \sum_k C_{ijk} {W_k(0)\o z^{i+j-k}}\,.}
We employ a unit normalization, $C_{ij0}=\delta_{ij}$. 

Explicit checks of the conjecture require a map between bases of the 6d half-BPS chiral ring and the $\W_N$ generators. In the so-called ``quadratic basis'' of $\W_N$, the structure constants are conjecturally completely determined \c{Prochazka:2014gqa}. A more physically natural basis for many purposes is the ``Virasoro primary basis'', in which the currents $W_k$ obey the Virasoro primary condition $L_{n>0}|W_k\ra=0$.\foot{This basis emerges naturally in the construction of $\W_N$ via Drinfeld-Sokolov reduction of $SL(N,\mathbb{R})$ and in the context of AdS$_3$ higher spin gravity \c{Campoleoni:2010zq}, and makes the triality symmetry of $\W_N$ manifest \c{Gaberdiel:2012ku}. We point out that, for the value of $c$ given above, the triality symmetry actually degenerates into a duality symmetry, $\W_N\simeq W_{-2N}$, where one should regard $c$ as fixed.} Many low-lying structure constants in the primary basis may be found in \c{Gaberdiel:2012ku,Prochazka:2014gqa}. In the $1/c$ expansion, there is a natural map between the single-trace half-BPS operators $S_k$, and the currents $W_k$ in the Virasoro primary basis:
\e{swmap}{S_k\quad \leftrightarrow \quad W_k}
At subleading orders in $1/c$, the 6d spectrum undergoes mixing between single-trace and multi-trace operators: for instance, $S_4$ mixes with the $\Dc[40]$ projection of :$S_2S_2$:. However, for $k=2,3$, no mixing is possible, due to the absence of $k=1$ operators in both the $\W_N$ algebra and the (2,0) spectrum.

While fundamentally a statement about the spectrum and OPE coefficients, the map to $\W_N$ has especially powerful implications for 6d four-point functions. As shown in \cite{Beem:2014kka}, one obtains a holomorphic function of $z$ from the 6d four-point function $\mathcal{G}_k(U,V;\sigma,\tau)$ by twisting $\s=\zb^{-2}$ and $\tau = (1-\zb^{-1})^2$. Given the map \eqr{swmap}, the claim of \cite{Beem:2014kka} is that
\es{2dG}{\Gc_k(z)|_{2d}&\equiv\mathcal{G}_k(z\bar z,(1-z)(1-\bar z);{\bar z}^{-2},(1-{\bar z}^{-1})^2)\,.\\
&= F_k(z)\,,
}
where $F_k(z)= z^{2k}\langle W_k(0) W_k(z) W_k(1) W_k(\i) \rangle $ is the four-point amplitude of identical spin-$k$ currents $W_k(z)$ of $\W_N$. Moreover, under the twist, $\Gc_k=\F_k$ in \eqr{Ward2}, so $\F_k(U,V;\s,\tau)$ is precisely the 6d uplift of the 2d correlator $F_k(z)$. To explicitly relate $\W_N$ structure constants $C_{ijk}$ to 6d OPE coefficients $\l_{ijk}$, we must determine how the chiral algebra twist relates the 6d blocks to 2d blocks. As derived in Appendix \ref{appblocktwist}, 
\es{2dblockExp}{
\mathcal{G}_k(z)\big\vert_{2d}=&\,1+\sum_{n=1}^k\lambda^2_{k,\mathcal{D}[2n\,0]}  \frac{4^n(1/2)_n}{(1)_n} g_{4n,0}(z)\\
&+\sum_{n=1}^{k-1}\sum_{j,=0,2,\dots}^\infty\lambda^2_{k,\mathcal{B}[2n\,0]_j}\frac{4^{n+1}(\frac12)_{n+1}}{(1)_{n+1}}A^{\mathcal{B}[2n\,0]}_{n+1\,n+1\,4n+6+j\,j+2}\times g_{4n+6+j,j+2}(z)\,,}
where the first term represents the unit operator, and 
\e{slt2B}{g_{\Delta,j}(z)=z^{\frac{\Delta+j}{2}}{}_2F_1\left(\frac{\Delta+j}{2},\frac{\Delta+j}{2},\Delta+j,z\right)}
are the $SL(2,\mathbb{R})$ global conformal blocks. The crossing equations for $F_k(z)$ imply that only $\lfloor {2k\o3}\rfloor$ of the infinitely many $\l^2_{k,\M_k}$ are independent \cite{Bouwknegt:1988sv, Headrick:2015gba}.
 
Of particular use in the following will be the OPE coefficients $\l^2_{k,\Dc[2n\,0]}$ for $k=2,3$. For $k=2$, this is nonzero only for $n=1$ (cf. \eqr{cTolam}), and the corresponding $\W_N$ OPE coefficient is $C_{222}\propto 1/c$, i.e. the cubic coupling of the stress tensor. On the other hand, $k\geq3$ contain non-trivial information to all orders in $1/c$. The 6d squared OPE coefficient $\lambda^2_{3,\mathcal{D}[40]}$ is determined by $(C_{334})^2$ of $\W_N$, with relative coefficient determined by the $n=2$ term in the first line of \eqref{2dblockExp}:
\e{OPE3}{
\lambda^2_{3,\mathcal{D}[40]}=\frac16(C_{334})^2 ~.
}
In unit normalization, $(C_{334})^2$ takes the rational form\cite{Hornfeck:1992he,Hornfeck:1993kp,Blumenhagen:1994wg,Gaberdiel:2012ku}
\e{c334}{(C_{334})^2 = \frac{144 (c+2) (N -3) (c (N +3)+2 (N-1) (4 N+3))}{c(5 c+22) (N-2) (c (N+2)+(N-1)(3N+2))}\,.}

Let us expand $\l^2_{3,\Dc[40]}$ in the large $c$ limit, using \eqr{c334} with the necessary identification $c=4N^3-3N-1$:
\e{334exp}{\l^2_{3,\Dc[40]} \approx \frac{24}{5c}-36\cdot 2^{\frac13}c^{-5/3}-\frac{2568}{25}c^{-2}-234\cdot 2^{\frac23}c^{-7/3}-588\cdot  2^{1\o3}c^{-8/3} + O(c^{-3})\,~.
} 
Starrting at $c^{-5/ 3}$, all powers of $c^{-{1/ 3}}$ are generated. Moreover, in the primary basis, all $\W_N$ structure constants $C_{k_1k_2k_3}$ with $k_i>2$ can be written as rational functions of $c$ and $C_{334}$ to some fractional power \c{Gaberdiel:2012ku,Prochazka:2014gqa,Linshaw:2017tvv}. This follows from the structure of the Jacobi identity, as recently proven in full generality in \c{Linshaw:2017tvv}.\foot{Unusual normalization conventions, such as the one in \c{Prochazka:2014gqa, Linshaw:2017tvv}, can lead to especially simple-looking structure constants. In any convention, $1/c$ power counting ensures that the normalized OPE coefficients scale as $\sim 1/\sqrt{c}$ to leading order. We also note that \c{Linshaw:2017tvv} proved the uniqueness of the $\W_N$ algebra, given the list of spectrum-generating currents.} For example, in our unit normalization,
\e{}{C_{444} = {3(c+3)\o c+2}C_{334} - {288(c+10)\o c(5c+22)}\,C_{334}^{-1}\,.}
The uplift to 6d then implies that {\it all} half-BPS OPE coefficients not involving $S_2$, the stress-tensor multiplet, have $1/c$ expansions of the form \eqr{lm123}. 

This is precisely the structure one expects based on the cubic coupling $\phi_1\phi_2\phi_3$ in the quantum effective action in AdS$_7$. In fact, this structure should be present in any M-theory compactification on AdS$\times \M$ for some compact transverse manifold $\M$. We assume $L_{\rm AdS}\simeq L_\M$. Dimensional reduction of the 11d action, $S_{11}$, on $\M$ generates an effective AdS$_{d+1}$ action, $S_{d+1}$. A key point is that, in general, non-perturbative AdS$_{d+1}$ amplitudes receive contributions from the descent of {\it all}-point amplitudes in 11d: the extra fields can have legs on $\M$.  If dimensional reduction of the $2n$-derivative action in 11d contributes as $g_{123}^{(2n)}$ to a cubic scalar AdS vertex $g_{123}\,\phi_1\phi_2\phi_3$ in the quantum effective action in AdS$_{d+1}$, then accounting for factors of $\lp$,
\e{g123}{g_{123} \sim {g_{123}^{(2)}\o c} + \sum_{n=4}^\i g_{123}^{(2n)}\,c^{-{7+2n\o 9}}\,.}
The form of \eqr{g123} follows from dimensional analysis, $\left(L_{\rm AdS}/ \ell_{11}\right)^9 \propto c$, combined with the fact that the reduction on $\M$ can only produce powers of $L_{\rm AdS}$, not $\ell_{11}$.\foot{Later, we will note an interesting consequence of relaxing the assumption that $\Lads\approx L_\M$.} To relate $g_{123}$ to the dual CFT OPE coefficient $\l_{k_1k_2k_3}$, one multiplies \eqr{g123} by a function 
with an infinite expansion in non-negative integer powers of $1/c \sim G_N$, which accounts for bulk loops. Specializing to the case $\M=S^4$, we equate the result with $\l_{k_1k_2k_3}^2$, thus inferring that $g_{123}^{(2n)}=0$ for $n\neq 4,7,10,\ldots$. The result is consistent with the minimal form $g_{123}^{(2n)}=0$ for $n\neq4,7$. As explained in the introduction, the latter is precisely compatible with the known structure of the M-theory action, thus furnishing compelling evidence for the $\W_N$ chiral algebra conjecture to all orders in $1/c$. Conversely, given the $1/c$ expansion of $\l_{k_1k_2k_3}^2$ in \eqr{lm123}, we have given a holographic argument for the structure of protected vertices in M-theory, in particular, the absence of 10- and 12-derivative terms.

\ssec{Mellin Amplitude}
\label{Mellin}

We will find it useful to express our four-point function in Mellin space. For this purpose it is useful to separate out the disconnected piece 
\es{disc}{
\mathcal{G}_\text{disc}=1+{U^{2k}\sigma^{k}}+\frac{U^{2k}}{V^{2k}}\tau^k\,.
}
The Mellin transform $M_k(s,t;\sigma,\tau)$ of the connected correlator $\mathcal{G}_\text{conn}\equiv\mathcal{G}-\mathcal{G}_\text{disc}$ is then:
\es{mellinDef}{
\mathcal{G}^\text{conn}_k(U,V;\sigma,\tau)=\int_{-i\infty}^{i\infty}\frac{ds\, dt}{(4\pi i)^2} U^{\frac s2}V^{\frac t2-2k}M_k(s,t;\sigma,\tau)\Gamma^2\left[2k-\frac s2\right]\Gamma^2\left[2k-\frac t2\right]\Gamma^2\left[2k-\frac u2\right]\,,
}
where the Mellin space variables $s$, $t$, and $u$ satisfy the constraint $s+t+u=8k$. The two integration contours run parallel to the imaginary axis, such that all poles of the Gamma functions are to one side of the contour. 

We can similarly define the Mellin transform $\widetilde{M}_k(s,t;\sigma,\tau)$ of the reduced correlator $\mathcal{H}_k(U,V;\sigma,\tau)$ defined in \eqref{Ward2} as \cite{Rastelli:2017ymc}
\es{mellinH}{
\mathcal{H}_k(U,V;\sigma,\tau)=\int_{-i\infty}^{i\infty}\frac{ds\, dt}{(4\pi i)^2} U^{\frac s2+1}V^{\frac t2-2k+1}\widetilde{M}_k(s,t;\sigma,\tau)\Gamma^2\left[2k-\frac s2\right]\Gamma^2\left[2k-\frac t2\right]\Gamma^2\left[2k-\frac {\tilde{u}}{2}\right]\,,
}
where $\tilde{u}=u-6$. This reduced Mellin amplitude $\widetilde{M}_k(s,t;\sigma,\tau)$ is related to the full Mellin amplitude ${M}_k(s,t;\sigma,\tau)$ by the Mellin space version of \eqref{Ward2}:
\es{MellinWard2}{
{M}_k(s,t;\sigma,\tau)=\widehat{\Theta}\circ\widetilde{{M}}_k(s,t;\sigma,\tau)\,,
}
where $\widehat{\Theta}$ is a complicated difference operator whose explicit form is given in Appendix \ref{SUSYWARD}, and should be thought of as the Mellin space version of the differential operator $\Upsilon$ in position space.\foot{As explained in \cite{Rastelli:2017ymc}, the Mellin transform of $\mathcal{F}_k$ can be consistently defined to be zero, and can be recovered as a subtle regularization effect in properly defining the contour integrals of the inverse Mellin transform.} The Mellin presentation of the superconformal Ward identity will make the physics of the flat space limit especially transparent.

\section{Holographic four-point functions at tree level}
\label{hol}

Let us now discuss the four-point correlator of the operators $S_{k}$ in the $A_{N-1}$ $(2,0)$ theory, with special emphasis on the cases $k=2,3$. We will compute the Mellin amplitudes allowed by the superconformal Ward identity with the constraint that no triple poles appear in the inverse Mellin transform \eqref{mellinDef}, which restricts us to tree-level amplitudes \c{Aharony:2016dwx,Rastelli:2017udc}. We will organize the solutions we find according to their maximal degree in the large $s,t,u$ limit. We will find $\lfloor\frac{2k}{3} \rfloor$ independent meromorphic  solutions, as well as an infinite tower of purely polynomial solutions with increasing degrees. We postpone a physical justification and interpretation of this expansion to the next section. Given knowledge of the single-trace spectrum of half-BPS operators, this section may be viewed as either an AdS or CFT calculation.

\subsection{Mellin amplitudes for $A_{N-1}$ $(2,0)$ theories}
\label{tree}


The main advantage of the Mellin space representation mentioned in the previous section is that in a theory with a holographic dual one can easily write down the {\em tree-level} expression for the connected part of the four-point function.  At tree level, the relevant Witten diagrams are contact diagrams and exchange diagrams, so the tree level Mellin amplitude for a correlator of four $S_k$'s is 
 \es{Mtree1}{
  M_{k,\text{tree}} =  M_{k,\text{$s$-exchange}} +M_{k,\text{$t$-exchange}}+M_{k,\text{$u$-exchange}}   + M_{k,\text{contact}}  \,,
 }
where the $t$- and $u$-channel exchange diagrams are related to the $s$-channel by crossing
\es{tandu}{
M_{k,\text{$t$-exchange}}(s,t;\sigma,\tau)&=\tau^k M_{k,\text{$s$-exchange}}(t,s;\sigma/\tau,1/\tau)\,,\\
M_{k,\text{$u$-exchange}}(s,t;\sigma,\tau)&=\sigma^k M_{k,\text{$s$-exchange}}(u,t;1/\sigma,\tau/\sigma)\,.
}
In Mellin space, the contact diagrams corresponding to vertices dressed with $n$ derivatives are order-$n$ polynomials in $s$, $t$, $u$. The $s$-channel exchange for a bulk field $\phi'$ dual to a boundary conformal primary operator $\O'$ of dimension $\Delta'$, traceless symmetric spin $j'$, and $\mathfrak{so}(5)$ irrep $[2b\,2(a-b)]$ is
\es{exchange}{
M^{\Delta',j',[2b\,2(a-b)]}_{k,\text{$s$-exchange}}(s,t;\sigma,\tau)=Y_{ab}(\sigma,\tau)\left[\sum_{m=0}^{m_\text{max}}\frac{\mathcal{Q}^{\Delta',2k}_{j',m}(t)}{s-(\Delta'-j'+2m)}+R_{j'-1}(s,t)\right]\,,
}
where $Y_{ab}(\sigma,\tau)$ is defined in \eqref{Yab3333}, $R_{j'-1}(s,t)$ is a degree $j'-1$ polynomial in $s$ and $t$, and the Mack polynomial $\mathcal{Q}^{\Delta',\Delta}_{j',m}(t)$ is a degree $j'$  polynomial in $t$ whose explicit form we give in Appendix \ref{Mack}. The meromorphic part in \eqref{exchange} is fixed to match that of the conformal block for the exchange of $\cO'$, which is required by conformal symmetry; its poles sit at the twists of $\Oc'$ and the twists $\D'-j'+2m$ of its conformal descendants. The truncation at $m_\text{max}=2k-(\Delta'-j')/2-1$ enforces the constraint that the poles not overlap with the Gamma function double poles in \eqref{mellinDef} that correspond to exchanges of double-trace operators $S_k \p^{2n}\p_{\mu_1}\ldots \p_{\mu_\ell}S_k$. This is required by the $1/N$ expansion \c{Aharony:2016dwx,Rastelli:2017udc}.

In our case, the scalar operators $S_k$ are dual to the bulk scalars $\phi_k$, which descend from linear combinations of the 11d graviton and three-form along $S^4$. These are the only elementary scalars in AdS$_7$.  As shown in \eqr{opemultEq}, $S_k\times S_k\supset \oplus_{n=2}^{2k}\Dc[n0]$ where $S_{k}$ is the bottom component of $\mathcal{D}{[k0]}$.\foot{The irreps $\Dc[n0]$ appearing here may be realized, as operators, both by single-trace superconformal primaries $S_n$, as well as multi-trace superconformal primaries. Only the former are elementary fields in AdS, and thus only these are the ingredients of Witten diagrams.} According to the standard GKPW dictionary, we expect an exchange diagram for each conformal primary operator in these multiplets.\foot{We note that the extremal multiplet $\mathcal{D}[2k\,0]$ and the operator with twist $4k$ in the $\mathcal{D}[2(k-1)\,0]$ multiplet do not contain any meromorphic parts, because their twists are large enough that $m_\text{max}<0$ in \eqref{exchange}. }
\begin{table}[htp]
\begin{center}
 \begin{tabular}{c|c|c|r}
multiplet&  dimension & spin & $\mathfrak{so}(5)$ irrep \\
  \hline
$\mathcal{D}{[20]}$&  $4$ & $0$ & ${\bf 14} = [20]$ \\
 & $5$ & $1$ & ${\bf 10} = [02]$ \\
 & $6$ & $2$ & ${\bf 1} = [00]$\\
   \hline
$\mathcal{D}{[40]}$&  $8$ & $0$ & ${\bf 55} = [40]$ \\
 & $9$ & $1$ & ${\bf 81} = [22]$ \\
 & $10$ & $0$ & ${\bf 35'} = [04]$\\
  & $10$ & $2$ & ${\bf 14} = [20]$\\
   & $11$ & $1$ & ${\bf 10} = [02]$\\
    & $12$ & $0$ & ${\bf 1} = [00]$\\
   \hline
$\mathcal{D}_{[60]}$&  $12$ & $0$ & ${\bf 140''} = [60]$ \\
  & $13$ & $1$ & ${\bf 260} = [42]$ \\
 & $14$ & $0$ & ${\bf 220} = [24]$\\
  & $14$ & $2$ & ${\bf 55} = [40]$\\
   & $15$ & $1$ & ${\bf 81} = [22]$\\
    & $16$ & $0$ & ${\bf 14} = [20]$\\
 \end{tabular}
\end{center}
\caption{The conformal primaries that appear in each half-BPS supermultiplet in $S_k\times S_k$ for $k=2,3$.}
\label{moreMult}
\end{table}%
The meromorphic part of the $s$-channel exchange diagram $\widehat{M}_{k,\text{s-exchange}}$ for a multiplet $\mathcal{D}{[n0]}$ can then be written (up to overall normalization) as a linear combination of the contributions from its conformal primary components, 
\es{exchange2}{
&\widehat{M}^{\mathcal{D}{[n0]}}_{k,\text{$s$-exchange} }=\sum_{(\Delta',j')\in\mathcal{D}[n0]} A^{\mathcal{D}[2n\,0]}_{nn\Delta'j'}\widehat{M}_{k,\text{$s$-exchange}}^{\Delta',j',[2n\,0]}\,,
}
where $\widehat{M}^{\Delta',j',[2n\,0]}_{k,\text{$s$-exchange}}$ denotes the meromorphic part of the exchange diagram as defined in \eqref{exchange}.  The relative coefficients $A^{\mathcal{D}[2n\,0]}_{nn\Delta'j'}$ can be fixed using the superconformal Ward identity in Appendix \ref{SUSYWARD}, and are the coefficients of the conformal bock contributions to the superconformal block as defined in \eqref{GExpansion}. For example, for $k=2,3$, we give the branching $\mathfrak{osp}(8^*|4)\mapsto  \mathfrak{so}(2)\oplus\mathfrak{so}(6)\oplus \mathfrak{so}(5)_R$ in Table \ref{moreMult}. The exchange amplitudes needed for these cases are
\es{exchange3}{
\widehat{M}^{\mathcal{D}{[20]}}_{k,\text{$s$-exchange} }=&~\widehat{M}_{k,\text{$s$-exchange}}^{4,0,[20]}-\frac15\widehat{M}_{k,\text{$s$-exchange}}^{5,1,[02]}+\frac{3}{175}\widehat{M}_{k,\text{$s$-exchange}}^{6,2,[00]}\,,\\
\widehat{M}^{\mathcal{D}{[40]}}_{k,\text{$s$-exchange} }=&~\widehat{M}_{k,\text{$s$-exchange}}^{8,0,[40]}-\frac29\widehat{M}_{k,\text{$s$-exchange}}^{9,1,[22]}+\frac{8}{189}\widehat{M}_{k,\text{$s$-exchange}}^{10,0,[04]}\\
+&~\frac{100}{6237}\widehat{M}_{k,\text{$s$-exchange}}^{10,2,[20]}-\frac{5}{1617}\widehat{M}_{k,\text{$s$-exchange}}^{11,1,[02]}\,.
}

The most general tree level Mellin amplitude may then be written as
\es{Mtree}{
M_{k,\text{tree}}=\sum_{n=1}^{k-1}\lambda^2_{k,\mathcal{D}[2n\,0]}\widehat{M}^{\mathcal{D}{[2n\,0]}}_{k,\text{exchange}}+({\rm polynomial})\,,
}
where $\lambda^2_{k,\mathcal{D}[2n\,0]}$ are the OPE coefficients squared defined in \eqref{SBDecomp}, $\widehat{M}_{k,\text{exchange}}=\widehat{M}_{k,\text{$s$-exchange}}+\widehat{M}_{k,\text{$t$-exchange}}+\widehat{M}_{k,\text{$u$-exchange}}$ as in \eqref{tandu}, and the polynomial term includes both the contact terms, which are polynomials of arbitrarily large degree in $s$ and $t$, as well as the polynomial $R_{j'-1}$ terms that appear in the full exchange diagrams \eqref{exchange}, which are at most linear in $s$ and $t$.

We will now plug the ansatz \eqref{Mtree} into the superconformal Ward identities. This will further constrain the solutions, which we organize by the maximal degree $p$ of the polynomial term. The solutions can be divided into those that have a meromorphic term and those that do not.

\subsection{Meromorphic solutions}

We first discuss those solutions that contain a meromorphic term that comes with a polynomial term of maximal degree $p$. By checking many cases we find that the most general ansatz is
\e{Merok}{
M^{(p)}_{k,\text{mero}}=a^{(p)}_{k,\mathcal{D}[\frac{4+2p}{3}\,0]} M^{(p)}_{k,\text{poly}}+\delta_{p,1}a^{(1)}_{k,\mathcal{D}[20]} \widehat{M}^{\mathcal{D}{[20]}}_{k,\text{exchange}}+\sum_{n=2}^{k-1} a^{(p)}_{k,\mathcal{D}[2n\,0]} \widehat{M}^{\mathcal{D}{[2n\,0]}}_{k,\text{exchange}}\,,
}
where $a^{(p)}_{k,\mathcal{D}[2n\,0]} $ are free coefficients, and $p=1,4,7\dots (3\lfloor {\frac{2k}{3}}\rfloor-2)$ for $n>1$. The fact that there are $k-1$ possible exchange terms (one for each $\mathcal{D}[2n\,0]$ with $1\leq n<k$) but only $\lfloor \frac{2k}{3}\rfloor$ meromorphic solutions follows from the same property of $\W_N$ correlators noted in Section \ref{chiral}; we will explain this further when we relate these solutions to CFT data in Section \ref{wardexp}. As we now show, the superconformal Ward identity relates the meromorphic terms to the polynomial piece $M^{(p)}_{k,\text{poly}}$.

We begin with $p=1$. As we will explain in the next section, these amplitudes descend from the 11d supergravity term, so we denote them by
\e{}{M^{(1)}_{k,\rm mero}\equiv M_{k,\,\rm sugra}\,.}
For all $k$, there is a solution of the form \eqr{Merok}, with all coefficients nonzero. For $k=2,3$ we find
\es{SUGRATree}{
M_{2,\,\rm sugra}= &a^{(1)}_{2,\mathcal{D}[20]}\left[ \widehat{M}^{\mathcal{D}{[20]}}_{2,\text{exchange}}+M_{2,\text{poly}}^{(1)}\right]\,,\\
M_{3,\,\rm sugra}= &a^{(1)}_{3,\mathcal{D}[20]}\left[ \widehat{M}^{\mathcal{D}{[20]}}_{3,\text{exchange}}+\frac{8}{15} \widehat{M}^{\mathcal{D}{[40]}}_{3,\text{exchange}}+M_{3,\text{poly}}^{(1)}\right]\,,\\
 }
where the polynomial terms are given in Appendix \ref{amps}. A more compact expression for these Mellin amplitudes is given by the reduced form $\widetilde{M}_k$ defined by \eqref{MellinWard2}, which for $k=2$ is 
\es{2222Sugra}{
\widetilde{M}_{2,\,\text{sugra}}(s,t)&=\frac{32a^{(1)}_{2,\mathcal{D}[20]}}{{(s-6)(s-4)(t-6)(t-4)(\tilde u-6)(\tilde u-4)}}\,,
}
and for $k=3$ is
\es{3333Sugra}{
\widetilde{M}_{3,\,\text{sugra}}(s,t)&=a^{(1)}_{3,\mathcal{D}[20]}\left[\widetilde{M}^{(1)}_{100}(s,t)+\sigma\widetilde{M}^{(1)}_{100}(\tilde u,t)+\tau\widetilde{M}^{(1)}_{100}( t,s)\right]\,,\\
\widetilde{M}_{100}^{(1)}&=\frac{32}{27}\frac{(s-7)}{(s-8)(s-6)(s-4)(t-10)(t-8)(\tilde u-10)(\tilde u-8)}\,.
}
Up to an overall normalization, these expressions match those in \cite{Rastelli:2017ymc}. 

Now take $p>1$. For $k>2$ -- and only for $k>2$ -- we find other meromorphic solutions to the Ward identity. The simplest example of this is $k=3$, where we find an additional meromorphic solution with $p=4$:
\e{R4Tree}{
 M_{3,\text{mero}}^{(4)} = a^{(4)}_{3,\mathcal{D}[40]}\left[ \widehat{M}^{\mathcal{D}{[40]}}_{3,\text{exchange}}+M_{3,\text{poly}}^{(4)}\right]\,,\\
 }
where the explicit form of $M_{3,\text{poly}}^{(4)}$ is given in \eqr{polyPart2}. Similar solutions exist for all $k>2$, where all admissible $a^{(4)}_{k,\mathcal{D}[2n\,0]}$ are nonzero for $n>1$. The form of these results, and the determination of $a^{(4)}_{3,\mathcal{D}[40]}$ from CFT, will be explained in Section \ref{wardexp}. 

\subsection{Polynomial solutions}
\label{polynomial}
We can also find purely polynomial solutions to the Ward identities. Note that the degree of these purely polynomial terms can in general be the same as that of the polynomial amplitudes $M_{k,\text{poly}}^{(p)}$ that come with the meromorphic solutions. If we define $\mathcal{N}_k(p)$ as the number of solutions of maximal degree $p$ for a given $k$, then the purely polynomial terms $ M_{k,\text{pure-poly}}$ in $M_{k,\text{tree}}$\footnote{These terms are a subset of the total number of polynomial terms in $M_{k,\text{tree}}$, because the meromorphic solutions come with polynomial terms.} take the form
 \e{GenSol}{
  M_{k,\text{pure-poly}} =  
  \sum_{p=4}^\infty \sum_{i = 1}^{\mathcal{N}_k(p) - \mathcal{N}_k(p-1)}  B_{k}^{(p, i)} M_{k,\text{poly}}^{(p, i)}\,,
 }
where $M_{k,\text{poly}}^{(p, i)}$ is a purely polynomial solution to the Ward identity of degree $p$ with coefficient $ B_{k}^{(p, i)}$, and $i$ counts the number of different polynomials with the same maximal degree. 

For $k=2$, the reduced Mellin amplitude for these solutions takes the simple form
\es{2222Hi}{
\widetilde{M}_{2,\text{poly}}^{(p,i)}=\frac{(s^2+t^2+u^2)^{a_i}(stu)^{b_i}}{{(s-6)(t-6)(\tilde u-6)}}\qquad \text{s.t.}\qquad 2a_i+3b_i\leq p-4\,,
}
where $a_i$ and $b_i$ are non-negative integers. The sum of the number of partitions of all positive integers $y\leq x$ into 2 and 3 is given by 
 \es{Gotn}{n(x) &=\sum_{y=0}^x\oint_{q=0}{q^{-y-1}\o (1-q^2)(1-q^3)}\\
 &= \Big\lfloor{\frac{6+(x+3)^2}{12}}\Big\rfloor\,,
 }
so the number $\mathcal{N}_2(p)$ of polynomial solutions of degree $p$ for $k=2$ is
 \es{num2}{
 \mathcal{N}_2(p)=n(p-4)\,.
 }

For $k=3$, the $\widetilde{M}_{3,\text{poly}}^{(p,i)}$ do not take such a simple form for all $p$, but we write the cases $p=5,6$ in Appendix \ref{amps}. In this case we find by checking many solutions that\footnote{This formula is equal to the number of solutions to the Ward identity in the flat space limit minus the one meromorphic solution with $p=4$. This naive counting of polynomial solutions does not work for higher $k$ though. }
 \e{num}{
 \mathcal{N}_3(p)=n(p-4)+n(p-5)+n(p-6)-1\,.
 }
 For higher $k$ we found no simple pattern for the number of polynomial solutions, but they can be easily computed case-by-case. We do, however, note the following feature: at $p=4$, for even $k$ only, there is a unique polynomial solution in addition to the unique meromorphic solution \eqr{Merok}. 
\sec{Uplifting to M-theory}\label{s5}

Having established the space of solutions to the superconformal Ward identity, we turn to their physical interpretation in the (2,0) CFT and the uplift to M-theory. This relies on the flat space limit of $\la S_kS_kS_kS_k\ra$, which we perform using an adaptation of Penedones' formula \c{Penedones:2010ue}. Our goals are twofold: first, to give a precise dictionary for how to recover 11d amplitudes in the $\lp\ll1$ expansion from these four-point functions; and second, to show on general grounds how the functional form of the 11d amplitude is reflected in, and may be inferred from, the properties of the 6d CFT correlators. In the next section we apply this technique to derive the $R^4$ contribution to the 11d graviton amplitude. 

\ssec{Flat space limit for arbitrary KK modes}

Let us first present the adaptation of Penedones' original formula to the AdS$_7\times S^4$ compactification \cite{Penedones:2010ue,Fitzpatrick:2011hu,Chester:2018aa}, for the Mellin amplitude $M_k$ corresponding to the four-point function of KK modes $S_k$. Denoting $L\equiv \Lads$, 
  \es{FlatLimit}{
\lim_{L\to \infty}L^3 \left({L/2}\right)^4V_4 \, M_k(L^2   s, L^2   t;\s,\tau) = \frac{1}{\Gamma(4k-3)} \int_0^\infty d \beta\, \beta^{4(k-1)} e^{-\beta} {\cal A}^{11}_{k} \left( {2 \beta}  s, {2 \beta}   t;\s,\tau \right)  \, ,
 }
where $ \left({L/2}\right)^4V_4=\frac{\pi^2}{6}L^4$ is the $S^4$ volume (required by dimensional analysis). We interpret ${\cal A}^{11}_{k}$ as the 11d flat spacetime amplitude of four supergravitons with momenta $k^\mu_i$ restricted to an $\R^7\simeq$ AdS$_7|_{L\rar\i}$, integrated against four supergraviton Kaluza-Klein mode wave functions on $S^4$ and contracted with $\mathfrak{so}(5)$ polarization vectors $Y_i$. We can write ${\cal A}^{11}_{k}$ explicitly as
\es{aintrel}{{\cal A}^{11}_{k}(s,t;\s,\tau) &= {Y_1^{(k)}Y_2^{(k)}Y_3^{(k)}Y_4^{(k)}\o (Y_1\cdot Y_2)^k(Y_3\cdot Y_4)^k}\sum_{\A,\B,\gamma,\del} {\cal A}^{11}_{\A\B\gamma\del}(s,t) \\&\times\,V_4 \int_{S^4} \!d^4x \sqrt{g}\, \Psi_{I_{11}\dots I_{1k}}^\A(x) \Psi_{I_{21}\dots I_{2k}}^\B(x) \Psi_{I_{31}\dots I_{3k}}^\gamma(x) \Psi_{I_{41}\dots I_{4k}}^\delta(x)\,,}
The ingredients are as follows: ${\cal A}^{11}_{\A\B\gamma\del}(s,t)$ is an invariant tensor in the supergraviton polarizations $\A,\B,\gamma,\del$; $\Psi_{I_{i1}\dots I_{ik}}^\A(x)$ is the normalized KK mode wave function for the particle $i$ on a unit $S^4$; and $Y_i^{(k)} =Y_i^{I_1}\cdots Y_i^{I_k}$ are the scalar $S^4$ harmonics. 

To actually extract $\Ac^{11}_k(s,t;\s,\tau) $ from the integral \eqr{aintrel} for arbitrary $k$ is not straightforward, nor is it necessary. On general grounds, the uplift to 11d must be proportional to the four-supergraviton amplitude, for any $k$. This follows from the fact that all operators $S_k$ are dual to KK modes of 11d supergravitons. Using this observation and matching the degree of the polarizations, the flat space limit of $M_k$ must yield the 11d amplitude in the form
\e{11damp}{\flim M_k(s,t;\s,\tau)  = \Ac^{11}_\perp(s,t;\s,\tau) P_{k-2}(\s,\tau)\,,}
where $P_{k-2}(\s,\tau)$ is a crossing-symmetric polynomial of degree-$(k-2)$ in $(\s,\tau)$, and 
$\Ac_\perp^{11}(s,t;\s,\tau)$ is defined as
\e{aka11}{(Y_1\cdot Y_2)^2(Y_3\cdot Y_4)^2\Ac_\perp^{11}(s,t;\s,\tau) = \Ac^{11}(p_i;Y_i)\big|_{p_i\cdot Y_i=0}}
The orthogonal kinematics $Y_i\cdot p_i=0$ follows from taking the flat space limit of amplitudes in a direct product spacetime like AdS$_7\times S^4$. Note that while $\Ac^{11}(p_i;Y_i)$ depends in general on the individual momenta, $\Ac_\perp^{11}(s,t;\s,\tau)$ only depends on Mandelstam invariants, as we demonstrate momentarily. 

\eqr{11damp} makes clear that $\flim M_k(s,t;\s,\tau)$ must reproduce the complete functional form of $\Ac^{11}$, which takes the form \eqr{fexp2}, subject to the orthogonal kinematics $p_i\cdot Y_i=0$. We now show that the kinematic factor $\widehat K$ in the 11d supergravity amplitude follows elegantly from the flat space limit of the 6d superconformal Ward identity \eqr{scwardflat}. By direct computation, one can show that
\es{scwardflat}{
\lim_{s,t\to\infty}M_k^{(p)}(s,t;\sigma,\tau) &\approx \left({4(k-1)+p\o 128}\right)\,\big(stu \,\Theta_4^{\flat}(s,t;\sigma,\tau)\big)\,\Mt_k^{(p)}(s,t;\s,\tau)|_{s,t\to\infty}\,,\\
\Theta_4^{\flat}(s,t;\sigma,\tau) &\equiv (tu+ts\s+su\tau)^2\,,
}
where in the flat space limit $u=-s-t$. The notation $\Theta_4^{\flat}$ refers to the fact that this is the flat space limit of the difference operator $\Theta_4$ defined in \eqref{theta6} that appears in the 4d $\mathcal{N}=4$ superconformal Ward identities \cite{Rastelli:2017udc} (where it is denoted by $\widehat R$). Turning now to $\widehat K$, this is equivalent to the $t_8t_8R^4$ tensor, where $R_{\mu\nu\rho\sigma}$ is the linearized Weyl curvature in momentum space. It may be defined as (e.g. \c{DHoker:2005jhf})
\e{}{\widehat K = ((m_1m_2)(m_3m_4) -  4(m_1m_2m_3m_4)+(\text{perms}))^2\,,}
where
\e{}{m_i^{\mu\nu} \equiv \z_i^{[\mu} p_i^{\nu]}~, \quad (m_im_j) \equiv m_i^{\mu\nu}m_j^{\nu\mu}~, \quad (m_im_jm_km_l) \equiv m_i^{\mu\nu}m_j^{\nu\rho}m_k^{\rho\sigma}m_l^{\sigma\mu} \,.}
$\z_i$ and $p_i$ are the polarization vector and momenta of the $i$'th 11d graviton, respectively. In general, $\widehat{K}$ is not just a function of $(s,t)$. But in 11d kinematics where $\z_i\cdot p_j=0$ for all $(i,j)$, one finds
\e{kkb}{\widehat K|_{\z_i \rar Y_i} = 4(Y_1\cdot Y_2)^2(Y_3\cdot Y_4)^2\Theta_4^{\flat}(s,t;\s,\tau)\,.}
Therefore, the universal factor $\Theta_4^{\flat}(s,t;\s,\tau)$ that is required by the superconformal Ward identity accounts for the overall momentum/polarization-dependent factor $\widehat K$ of the 11d graviton amplitude.\foot{In \cite{Chester:2018aa}, the appearance of $\Theta_4^{\flat}(s,t;\s,\tau)$ in the four-point functions of stress tensor multiplets in ABJM was derived by appealing to maximal gauged supergravity in AdS$_4$.}  It is satisfying that 6d superconformal symmetry generates the $\widehat K$ factor in the uplift for any $k$. As noted in the introduction, the fact that $\Theta_4$ also appears in the 4d $\mathcal{N}=4$ superconformal Ward identities, combined with \eqr{kkb}, implies that type IIB string amplitudes are proportional to $\widehat K$ to all loop orders.\foot{The overall factor in type IIA and IIB string theory can in principle differ at five loops and beyond \c{Berkovits:2006vc}, but is sometimes indirectly argued to be equivalent to all orders (e.g. \c{Green:2008bf}). We thank Oliver Schlotterer for a discussion on this point.} 

Returning to \eqr{11damp}, we note that in a ratio of amplitudes at different orders in $\lp$, $P_{k-2}(\s,\tau)$ will cancel. Therefore, given the form of \eqr{fexp2}, we may express the tree-level terms $f_{D^{2m}R^4}(s,t)$ in terms of the basis of polynomial and meromorphic solutions of fixed degree $p=4+m$:
\es{fExp}{
&f_{D^{2m}R^4}(s,t)=  \frac{1}{2^{m+3}(4k-2)_{m+3}}\lim_{s,t\to\infty} \left[ \frac{\sum'_iB_{k}^{(4+m,i)}M^{(4+m,i)}_{k,\text{poly}}(s,t;\sigma,\tau)}{M_{k,\rm sugra}(s,t;\sigma,\tau)}\right.\\
&\left. ~~~~~~~~~~~~~~~~~~~~~~~~~~~+\begin{cases}\frac{a^{(4+m)}_{k,\mathcal{D}[4+2m/3\,0]}{M}^{(4+m)}_{k,\text{mero}}(s,t;\sigma,\tau)}{M_{k,\rm sugra}(s,t;\sigma,\tau)}\quad \text{for}\quad \frac{2k-1}{6}\geq m\in 3\mathbb{Z}\\
0\quad\text{otherwise}\end{cases}\right]\,,
}
where the numerical prefactor comes from the $\b$-integral in \eqr{FlatLimit}. This is one of our main formulas. 

An important point is that the sum $\sum'_i$ in \eqr{fExp} is defined to run only over all Mellin amplitudes whose $(\sigma,\tau)$-dependence is given by $P_{k-2}(\s,\tau)$. This polynomial is not a function of $\lp$, so it can be computed once and for all from, say, taking the flat space limit of the supergravity term $M_{k,\text{sugra}}(s,t;\sigma,\tau)$. This places strong constraints on the 6d Mellin amplitudes, not all of which have this factorized form. For instance, for $k=3$, the polynomial is unique up to rescaling
\e{f1z}{P_1(\s,\tau) =1+\s+\tau\,.}
On the other hand, there are many $k=3$ solutions of the 6d superconformal Ward identity that do not have this structure: at $p=7$, for example, we find
\es{3333large}{
 \lim_{s,t\to\infty}{M}_{3,\text{poly}}^{(7,1)}&\propto\Theta_4^{\flat}(s,t;\s,\tau)(1+\sigma+\tau)stu\,,\\ \lim_{s,t\to\infty}{M}_{3,\text{poly}}^{(7,2)}&\propto\Theta_4^{\flat}(s,t;\s,\tau)(s+u\sigma+t\tau)(s^2+t^2+u^2)\,.
}
${M}_{3,\text{poly}}^{(7,2)}$ thus must not appear in the 6d amplitude at $O(c^{-7/3})$. For $k=4$, 
\es{}{ P_2(\s,\tau)= (1+\s^2+\t^2) + b(\s+\t+\s\t)}
for some constant $b$, which can be determined from supergravity \c{Rastelli:2017ymc} to be $b=4$. For a given $k$, there are $n(k-2)$ independent orbits of crossing, where $n(x)$ was introduced in  \eqr{Gotn}, one linear combination of which is picked out by M-theory.\foot{One can check that $n(k-2)$ is equal to, but simpler than, $\Nc_{k-2}$ as defined in eq. 4.43 of \c{Rastelli:2017udc}. This simplification is the result of trying.} We can state the general criterion for which solutions can contribute to $f_{D^{2m}R^4}(s,t)$ in terms of the $\Mt_{lmn}$ (cf. \eqr{mtlmn}): they must be crossing-symmetric at large $s,t$. This discussion makes clear that 11d superPoincare invariance is more constraining that the flat space limit of the 6d superconformal Ward identity.

\ssec{Explaining the momentum expansion of Mellin amplitudes}\label{wardexp}
Now that we can perform the flat space limit, we return to the previous section's mathematical solutions to the superconformal Ward identities, and interpret them as solutions of the actual (2,0) CFT. The main point is that the degree of the Mellin amplitudes is correlated with powers of $1/c\sim \ell_{11}^9$. This is visible from the flat space limit (cf. \eqr{fExp}), which determines the corresponding power of momenta, and hence of $\lp$, in the corresponding 11d S-matrix element. In particular, the Mellin amplitude coefficients $ B_{k}^{(p, i)}$ and $a^{(p)}_{k,\mathcal{D}[2n\,0]}$ obey
\es{cExp}{a^{(p)}_{k,\mathcal{D}[2n\,0]}\,,~B_{k}^{(p, i)}\sim c^{-\frac{2p+7}{9}}(1+O(c^{-{2/9}}))\,.}
Given \eqr{cExp} and the form \eqr{Mtree} of the tree-level amplitudes, we see that the coefficients $a^{(p)}_{k,\mathcal{D}[2n\,0]}$ are the squared OPE coefficients for operators in $\Dc[2n\,0]$ multiplets, evaluated at $O(c^{-\frac{2p+7}{9}})$:
\e{alambda}{\l^2_{k,\Dc[2n\,0]} = \sum_{p=1}^\i c^{-{2p+7\o9}}a^{(p)}_{k,\mathcal{D}[2n\,0]}\,.}
%
This explains why we found only $\lfloor \frac{2k}{3}\rfloor$ meromorphic Mellin amplitudes when there are $k-1$ different supermultiplets $\mathcal{D}[2n\,0]$ exchanged: as recalled in Section \ref{chiral}, holomorphy and crossing in $\W_N$ completely determine the four-point function in terms of only $\lfloor \frac{2k}{3}\rfloor$ independent OPE coefficients. The fact that for $n=1$ we have just a single coefficient $a^{(1)}_{k,\mathcal{D}[20]}$ is explained by the fact that $\l^2_{k,\Dc[20]}\propto c^{-1}$ (cf. \eqr{cTolam}). At $p=1$, where there exists a meromorphic solution for all $k$ with $a^{(1)}_{k,\Dc[2n0]}\neq 0$ for all $n$, we see that this is just the two-derivative amplitude in AdS$_7$, which descends from 11d supergravity; this explains the $M_{k,\rm sugra}$ notation introduced earlier. At $p=4$, the existence of meromorphic solutions for all $k>2$ reflects the fact that the three-point functions $\l^2_{k,\Dc[2n\,0]}$ for $n>1$ receive corrections of $O(c^{-5/3})$ which descend from the 11d $R^4$ term (+ superpartners).\foot{This is the same mechanism that generates an $R^3$ term in AdS$_7$ which is responsible for the $O(N)$ contribution to the (2,0) central charge \c{Tseytlin:2000sf}.} This is indeed visible in the CFT where $\l^2_{k,\Dc[2n\,0]}$ are equivalent to $\W_N$ structure constants, as discussed around \eqr{334exp}--\eqr{g123}.

\sssec{Dimensional reduction and M-theory constraints on crossing}\label{crosm}

The scaling \eqr{cExp} may also be seen using dimensional reduction of M-theory.\footnote{To recover the exact coefficient of each amplitude from dimensional reduction would require knowledge of the full supersymmetric completion of the 11d higher derivative terms, which is unknown aside from the $R^4$ term. The action analysis may be seen as a book-keeping device for the $c$-scaling of on-shell amplitudes, which are what we actually compute.} On general grounds, the quartic part of the effective Lagrangian in AdS$_7$ is constrained to take the form
\e{ldq}{\Lc_{7}^{\rm quartic} = c^{-2/3}R^4f_{R^4}(c)+ c^{-8/9}D^2R^4f_{D^2R^4}(c)+c^{-10/9}D^4R^4f_{D^4R^4}(c)+\ldots}
where we have used $D^{2m}R^4$ to denote all $(8+2m)$-derivative terms in the 7d action, such as $(\p^2\phi_3)^4$. The functions $f_i(c)$ have $1/c$ expansions 
\e{ldq2}{f_i(c) = \sum_{n=0}^\i f_i^{(n)} c^{-{2n\o 9}}~, \quad f_i^{(n)}\in\R}
\eqr{ldq} follows from the same arguments as in Section \ref{chiral}: dimensional reduction correlates the power of $1/c$ with the number of 11d derivatives via \eqr{LellpRelation}. The constants $f_i^{(n>0)}$ descend from dimensional reduction of terms in 11d with with legs on $S^4$, i.e. from 11d terms with $2n$ {\it more} derivatives than the AdS$_{7}$ vertex.\foot{New tensor structures can also appear after the dimensional reduction that are not present in 11d. For instance, $(R_{\mu\nu}R^{\mu\nu})^3$ in 11d can generate $(R_{\mu\nu}R^{\mu\nu})^2$ in AdS, which is different from the $t_8t_8R^4$ and $\eps_{11}\eps_{11}R^4$ tensors that appear in 11d.} A similar argument applies to cubic vertices in AdS$_7$ which give rise to meromorphic Mellin amplitudes: these also admit an expansion in fractional powers of $1/c$, as observed in \eqr{g123} using $\W_N$. Given these facts, the relation \c{Heemskerk:2009pn} between bulk derivatives and polynomial solutions to crossing symmetry implies \eqr{cExp} to leading order in $1/c$. The above discussion applies without modification to any AdS$\times \M$ spacetime of M-theory with $\Lads \approx L_\M$.

The above insights have consequences for solutions to four-point crossing in any CFT dual to an M-theory AdS$\times \M$ compactification (with $\Lads \approx L_\M$), which go beyond the implications of the flat space limit. At $O(c^{-(7+2p_{\rm max})/9})$, the allowed solutions to crossing are those with $p\leq p_{\rm max}$ (see \eqr{cExp} and \eqr{GenSol}); in fact, the polynomials with $p < p_{\rm max}$ -- which do not survive the flat space limit -- are precisely dual to terms $f_i^{(n>0)}$ in \eqr{ldq}--\eqr{ldq2}. It is long-known that in M-theory, there are no 10- and 12-derivative terms. Therefore, there are no solutions to crossing at orders $c^{-17/9}$ and $c^{-19/9}$. Applying this to $\M=S^4$, together with the results of the next section for the $k=2,3$ four-point functions in the (2,0) CFT through $O(c^{-5/3})$, we have explicitly determined these correlators at tree-level up to $O(c^{-7/3})$. This rules out some of the low-lying solutions of \c{Heslop:2017sco}.\foot{It has been a long-standing goal in holography to find explicit examples of AdS $\times \M$ compactifications with a parametric hierarchy $L_{\rm AdS} \gg L_\M$. Such CFTs would have an especially sparse spectrum of light, low-spin single-trace operators. In this case, dimensional reduction will generate positive powers of $\ell_{11}/L_\M\equiv c_\M^{-1/9}\gg c^{-1/9}$. The quartic effective action in AdS will again take the form \eqr{ldq}, but where $f_i(c) \rar\ f_i(c_\M)$. Relating this to solutions to CFT crossing gives a new diagnostic, using CFT four-point functions, of whether a large $c$ CFT has an M-theory dual with $L_\M\gg L_{\rm AdS}$: at a given order in $1/c$, the {\it only} polynomial solutions to crossing are those of maximal degree $p=p_{\rm max}$. It would be interesting to use this in an effort to bootstrap the existence of such CFTs.}

\section{M-Theory from CFT Data}
\label{compare}

With all pieces in place, we now relate M-theory in the $\lp\ll1$ expansion to CFT data. We first derive $R^4$ from CFT. We then lay the groundwork for deriving the tree level higher derivative terms $\sim D^{2m}R^4$ from CFT. 

\subsection{$R^4$ from $\langle S_3S_3S_3S_3\rangle$}
\label{R4extract}
For the case $k=3$, the $\W_N$ algebra gives us a single nontrivial constraint from the OPE coefficient $\lambda_{3,\mathcal{D}[40]}$, which is enough to fix $M_{3,\text{mero}}^{(4)}$. We gave the $R^4$ contribution to $\Ac^{11}$ in \eqr{fr4}, whose derivation we review in Appendix \ref{appstring}.
%
%
Plugging it into our formula \eqr{fExp} with $m=0$ and $k=3$, we obtain the prediction
\es{fExp23}{\frac{a^{(4)}_{3,\mathcal{D}[40]}}{a^{(1)}_{3,\mathcal{D}[20]}}=-32\left(\frac{\ell_{11}}{\Lads}\right)^6\,.}
Translating to CFT data using \eqr{LellpRelation} and \eqr{alambda},
\es{fExp232}{{\l^2_{3,\Dc[40]}\big|_{c^{-5/3}}\o \l^2_{3,\Dc[20]}\big|_{c^{-1}}}=-4\cdot 2^{\frac13}c^{-\frac23}\,.}
This precisely matches the OPE coefficients \eqr{cTolam} and \eqr{334exp} as derived from the CFT with help from the $\W_N$ chiral algebra conjecture. Thus, we have derived the $R^4$ coefficient from 6d CFT data. We point out that $\l^2_{3,\Dc[40]}\big|_{c^{-5/3}} \sim c^{-1}N^{-2}$, and thus is, fortunately, independent of the $O(N)$ term in $c$ whose 11d origin is $R^4$ itself \c{Tseytlin:2000sf}. 

\subsection{Higher derivatives from $\langle S_2S_2S_2S_2\rangle$}
\label{higherMatch}

For higher order terms $\sim D^{2m}R^4$, $\W_N$ does not provide enough constraints on the $k=3$ amplitude to completely fix the solutions, due to the existence of pure polynomial solutions. Instead, our strategy will be to relate the higher-degree Mellin amplitude coefficients $B^{(p,i)}$ to CFT data. We will focus on the lowest case $k=2$, in which case the higher-derivative Mellin amplitudes are the purely polynomial amplitudes $M_{2,\text{poly}}^{(p, i)}$ of degree $p$, which are defined using the Mellin space Ward identity \eqr{MellinWard2} and the reduced Mellin amplitudes $\widetilde{M}_{2,\text{poly}}^{(p, i)}$ in \eqref{2222Hi}. We will restrict to $p\leq10$ where we can extract unambiguous information from tree level Mellin amplitudes, without contamination from loop-level data.\foot{The 2-loop 11d amplitude first appears at $p=10$, which makes it impossible to fix $M_k^{(p, i)}$ for $p>11$ purely in terms of tree level data. For $p=10$, while the term that scales like $c^{-3}$ will receive contributions from loop amplitudes, there is also a $c^{-3} \log c$ term that is fixed by the logarithmic divergence of the 2-loop amplitude in 11d supergravity; this should be fixable using tree level CFT data and the techniques of \c{Aharony:2016dwx, Alday:2017vkk}.}

To extract CFT data from the purely polynomial Mellin amplitudes $M_{2,\text{poly}}^{(p,i)}$ in terms of their coefficents $B_2^{(p,i)}$ for $p=6,7,8,9,10$ (where $i=1$ except for $p=10$ where $i=1,2$), we will use the algorithm developed for extracting CFT data from Mellin amplitudes in 3d \cite{Chester:2018lbz}. We use the following normalization for the conformal blocks in the lightcone limit $U\to0$, fixed $V$:
\es{lightLimit}{
\lim_{U\to0}G_{\Delta,j}(U,V)=U^{\frac{\Delta-j}{2}}(1-V)^j \,{}_2F_1\left(\frac{\Delta+j}{2},\frac{\Delta+j}{2},\Delta+j,1-V\right)
}
This calculation is very similar to that of \cite{Chester:2018lbz,Chester:2018aa}, so we will only briefly sketch the derivation. 

The amplitudes $M_2^{(p,i)}$ will contribute to the anomalous dimension of the infinite tower of unprotected double-trace conformal primary operators
\e{}{[S_2S_2]_{n,j} \simeq S_2\p^{2n}\p_{\mu_1}\ldots \p_{\mu_j}S_2-({\rm traces})\,.}
As discussed in \cite{Heemskerk:2009pn,Heslop:2017sco,Alday:2014tsa}, a purely polynomial Mellin amplitude of maximal degree $p$, which corresponds to a flat space vertex with $2p$ derivatives, contributes to the double-trace operators with spin $j\leq p-4$. We will now show how to fix the $\mathcal{N}(p) - \mathcal{N}(p-1) $ coefficients $B_{2,(p,i)}$, indexed by $i$, of each degree $p$ tree level term $M^{(p,i)}_{2,\text{poly}}$ defined by acting with \eqref{theta6} on \eqref{2222Hi} by extracting at least $\mathcal{N}_2(p) - \mathcal{N}_2(p-1) $ different pieces of CFT data from these amplitudes. We will use the OPE coefficients squared $a_\mathcal{M}$ of the protected multiplets $\mathcal{M}\in\{\mathcal{D}[04],\mathcal{B}[02]_j\}$ from \eqref{opemultEq} that are not fixed by $\W_N$, as well as the scaling dimension of the lowest twist long multiplet with spin $j$. The supergravity contribution to these quantities ($\sim c^{-1}$) was computed in \cite{Arutyunov:2002ff,Heslop:2004du}. The higher derivative Mellin amplitudes $M^{(p,i)}_{k,\text{poly}}$ discussed above will contribute starting at order $c^{-\frac{7+2p}{9}}$, and then will generically include all subleading powers of $c^{-2/9}\sim \lp^2$ (see \eqr{ldq}).

Let's define the quantity $\g^{(p,i)}_{\mathcal{A}[00]_{j+8,j}}$ as the contribution of $M_2^{(p,i)}$ to the anomalous dimension $\g_{\mathcal{A}[00]_{j+8,j}}$ of the leading twist operators $[S_2S_2]_{0,j}$. We focus only on leading twist for simplicity, because higher twists are degenerate. We will use the conformal primary $(j+12,j)_{[40]}$, because it is the only conformal primary in that $R$-symmetry channel, so we do not have to worry about mixing with other conformal primaries. To extract these we will need the product of the mean field theory (MFT) OPE coefficient squared $a^{\rm MFT}_{\mathcal{A}[00]_{j+8,j}}$ and the coefficient $A_{22\,j+12\, j}^{\mathcal{A}[00]}(j+12,j)$, which as shown in \cite{Heslop:2004du} in our conventions are
\es{OPEd6}{
A_{22\,j+12\, j}^{\mathcal{A}[00]}(j+12,j)a^{\rm MFT}_{\mathcal{A}[00]_{j+8,j}}=\frac{(j+1)(j+2)(j+9)(j+10)(j+5)!(j+6)!}{360(9+2j)!}\,.
}
Using these quantities and following the algorithm in \cite{Chester:2018aa}, we find the results listed in the last four rows of Table \ref{resultList}. 

For the protected OPE coefficients, we only need to worry about mixing with multiplets that are not in the chiral algebra, because those in the chiral algebra do not receive corrections beyond supergravity. For $a_{\mathcal{D}[04]}$, we can see from the tables of supermultiplets in Appendix \ref{super20} that its superconformal primary does not appear in any other supermultiplets, so we can easily extract its OPE coefficient. For $a_{\mathcal{B}[02]_j}$, the superconformal primary now also appears in $\mathcal{D}[04]$, so for simplicity we will extract its OPE coefficient from the conformal primary $(j+11,j+1)_{[40]}$. Using the superblock coefficients computed in Appendix \ref{super20}, we find the results listed in the first four rows of Table \ref{resultList}.

\begin{table}[htp]
\begin{center}
\begin {tabular} {| c || c | c | c | c | c | c | c |}
\hline
 {CFT data:}&$M_2^{(4,1)}$  &$M_2^{(6,1)}$ & $M_2^{(7,1)}$ &$M_2^{(8,1)}$ &$M_2^{(9,1)}$ &$M_2^{(10,1)}$ &$M_2^{(10,2)}$ \\
  \hline
\TBstrut  $a_{\mathcal{D}[04]}$&  $-\frac{3}{7}$& $-\frac{1308}{77}$ & $-\frac{1224}{77}$ & $-\frac{692304}{1001}$& $-\frac{575712}{1001}$ & $-\frac{28496064}{1001}$ & $-\frac{755136}{1001}$ \\
  \hline
 \TBstrut  $a_{\mathcal{B}[02]_1}$& 0& $-\frac{2000}{1573}$ & $\frac{6000}{1573}$ & $-\frac{115200}{1573}$& $\frac{33600}{1573}$ & $-\frac{103148800}{26741}$ & $\frac{835200}{1573}$ \\
  \hline
 \TBstrut  $a_{\mathcal{B}[02]_3}$& 0& $0$ & $0$ & $-\frac{1354752}{158015}$& $\frac{4064256}{158015}$ &-$\frac{1175924736}{3002285}$ & $-\frac{12192768}{158015}$ \\
  \hline
\TBstrut  $a_{\mathcal{B}[02]_5}$& 0& $0$ & $0$ & $0$& $0$ & $-\frac{89579520}{1356277}$ & $0$ \\
  \hline
 \TBstrut $\g_{\mathcal{A}[00]_{8,0}}$& $-\frac{250}{539}$ & $-\frac{197000}{7007}$ & $-\frac{96000}{7007}$ & $-\frac{12148000}{7007}$& $-\frac{384000}{539}$ & $-\frac{12853840000}{119119}$ & $-\frac{6144000}{7007}$ \\
  \hline
 \TBstrut   $\g_{\mathcal{A}[00]_{10,2}}$& 0& $-\frac{75264}{24167}$ & $\frac{301056}{24167}$ & $-\frac{122228736}{410839}$& $\frac{10838016}{31603}$ & $-\frac{184945926144}{7805941}$ & $\frac{828506112}{410839}$ \\
  \hline
 \TBstrut     $\g_{\mathcal{A}[00]_{12,4}}$& 0& $0$& $0$ & $-\frac{1280240640}{146834831}$& $\frac{5120962560}{146834831}$ & $-\frac{9510359040}{11294987}$ & $-\frac{20483850240}{146834831}$ \\
  \hline
 \TBstrut    $\g_{\mathcal{A}[00]_{16,6}}$& 0& $0$& $0$ & $0$& $0$ & $-\frac{2569273344000}{130985163829}$ & $0$ \\
  \hline
\end{tabular}
\end{center}
\caption{Contributions to the OPE coefficients squared $a$ and anomalous dimensions $\g$ of various multiplets appearing in $S_2\times S_2$ from the degree $p$ polynomial Mellin amplitudes $M^{{(p,i)}}_{2,\text{poly}}$. The latter are defined via \eqr{MellinWard2} and the reduced Mellin amplitudes $\widetilde M^{{(p,i)}}_{2,\text{poly}}$ in \eqref{2222Hi}. The entries should be understood as multiplying $B_2^{(p,i)}$, the overall coefficient of $M^{{(p,i)}}_{2,\text{poly}}$. Upon fixing the $B_2^{(p,i)}$ by comparing to an independent CFT computation of a quantity in the left-hand column, one determines the M-theory amplitude via \eqr{fExp}.
}\label{resultList}
\end{table}

\section{Conclusion}
\label{conc}

This paper developed an explicit program for how to extract the perturbative expansion of the 11d flat space S-matrix from the OPE data of the 6d $A_{N-1}$ $(2,0)$ CFT. We mostly focused on the flat space limit of the Mellin amplitude of four bottom components $S_k$ of the half-BPS multiplet $\mathcal{D}[k0]$, in an expansion at large central charge. In particular, we computed the $R^4$ term in 11d by using the protected 2d chiral algebra of the 6d theory, the quantum $\W_N$ algebra, to fix the four point function $\langle S_3S_3S_3S_3\rangle$ to $O(c^{-5/3})$, where $\mathcal{D}[30]$ is the lowest multiplet above the stress tensor multiplet. This computation relied crucially on the detailed form of the $\W_N$ algebra. Via their map to protected three-point functions in 6d, the $\W_N$ structure constants in the $1/c$ expansion were also shown to exhibit the absence of 10- and 12-derivative terms in the 11d effective action. Altogether, the aforementioned matches to 11d physics provide strong support for the chiral algebra conjecture of \c{Beem:2014kka}. Moreover, we provided an explicit roadmap for how the first several low-lying higher-derivative tree level terms $\sim D^{2m}R^4$ in the 11d S-matrix, including unknown terms beyond $D^6R^4$, can be directly recovered from 6d CFT data that is as yet unknown.

Our results give a new motivation for computing 6d CFT data. The only known method at this time of computing unprotected 6d CFT data is the numerical conformal bootstrap. This program was initiated in \cite{Beem:2015aoa} for $\langle S_2S_2S_2S_2\rangle$, but the present bounds are not strong enough to extract the $1/c$ expansion necessary to determine the M-theory effective action via the method of Section \ref{compare}. One lesson from our paper is that $\langle S_3S_3S_3S_3\rangle$ can be more constraining (and constrained) than $\langle S_2S_2S_2S_2\rangle$: the $\W_N$ chiral algebra contributes terms with a nontrivial expansion in $1/c$ to the former, but not the latter. By applying the numerical bootstrap to $\langle S_3S_3S_3S_3\rangle$, one could hope to find strong bounds on CFT data. For certain protected operators, these bounds could be compared to the nontrivial functions of $c$ determined from the chiral algebra. If these analytic functions were to saturate the numerical bounds, then one could use the extremal functional method \cite{ElShowk:2012hu} to read off the CFT data of all operators that appear in the four-point function, as was initiated in the case of the 3d maximally-supersymmetric ABJM theory in \cite{Agmon:2017xes}. One could also consider using the exact $\W_N$ result \eqr{c334} as input to this computation, which would presumably generate stronger bounds for the remaining OPE data.  

There are also several details of 6d Mellin amplitudes for $\langle S_kS_kS_kS_k\rangle$ that we would like to understand better. For instance, while we determined the number of purely polynomial solutions to the superconformal Ward identities for $k=2,3$, we were unable to find a simple pattern for larger $k$. For $k\geq4$, the operators $S_k$ undergo mixing with multi-trace operators in the respective $\Dc[2k\,0]$ R-symmetry representations, which might explain the counting in these cases. We would also like to better understand the relation between subleading $1/c$ corrections to a Mellin amplitude and terms in the $11d$ effective action. As also noted in \cite{Chester:2018aa}, these correspond to local higher-point vertices in 11d that involve more than four fields, e.g. $R^7$. Thus, the ``finite size corrections'' to the flat space limit of CFT correlators may be understood in part\footnote{Through 14-derivative order, supersymmetry relates $D^{2m}R^4$ to $R^{4+m}$. Starting at 16 derivatives, the uplift of the subleading terms that descend from $D^{2m}R^4$ in 11d cannot be related in any known way to higher-point terms.} as suitable soft limits of 11d higher-point amplitudes with four external gravitons. It would be interesting to make this relationship explicit.

Lastly, it would be interesting to extend the methods here and in \c{Chester:2018aa} to CFTs with semiclassical string theory duals, such as the large $N$ `t Hooft limits of $\mathcal{N}=6$ ABJM in 3d or $\mathcal{N}=4$ SYM in 4d. The complete string moduli dependence of the $D^8R^4$ term in type IIA and IIB is unknown despite many years of sophisticated direct attempts (e.g. \c{Green:2005ba,Green:2006gt,Bjornsson:2010wm,Vanhove:2010nf,Wang:2015aua} and references therein).\foot{We would be remiss not to highlight the recent work \cite{Bern:2018jmv}, which sheds a (negative) light on the status of possible non-renormalization of $D^8R^4$ by a direct five-loop supergravity computation.} It would be fascinating to determine this using holography. 

\section*{Acknowledgments} 

We thank Ofer Aharony, Chris Beem, Clay Cordova, Matthias Gaberdiel, Michael B. Green, Ken Intriligator, Silviu Pufu, Leonardo Rastelli, Oliver Schlotterer, Yifan Wang, Xi Yin and Xinan Zhou for useful discussions. We also thank Silviu Pufu and Leonardo Rastelli for comments on the draft. SMC is supported in part by the Simons Foundation Grant No~488651 and the Bershadsky Family Scholarship in Science or Engineering. EP is supported by Simons Foundation grant 488657, and by the Walter Burke Institute for Theoretical Physics. 

\appendix

\section{Superconformal Ward identity and $\mathfrak{so}(5)$ harmonics}
\label{SUSYWARD}

In position space, the superconformal Ward identity takes the form
\es{3D}{
\left(z\partial_z-2\alpha\partial_\alpha\right)\mathcal{G}\big\vert_{\alpha=z^{-1}}=\left(\bar z\partial_{\bar z}-2\alpha\partial_{\alpha}\right)\mathcal{G}\big\vert_{\alpha=\bar z^{-1}}=0\,,\\
}
where we define
 \es{uvsigmatauDefsz}{
  U \equiv z\bar z\,,\qquad V \equiv (1-z)(1-\bar z) \,, \qquad \sigma\equiv \alpha\bar\alpha\,,\qquad \tau\equiv(1-\alpha)(1-\bar\alpha) \,.
 }
 To implement the Ward identities in Mellin space, we first expand ${\cal G}_k(U,V;\sigma,\tau)$ into the R-symmetry polynomials $Y_{ab}(\sigma,\tau)$ as
\ie
{\cal G}_k(U,V;\sigma,\tau) = \sum_{a=0}^k\sum_{b=0}^a Y_{ab}(\sigma,\tau){\cal G}_{k,ab }(U,V) \,,
\fe
which has Mellin transform \eqref{mellinDef}
\es{mellinWard}{
{M}_k(s,t;\sigma,\tau) = \sum_{a=0}^k\sum_{b=0}^a Y_{ab}(\sigma,\tau){M}_{k,ab }(s,t) \,.
}
For $k=2,3$, the explicit forms of the $Y_{ab}(\sigma,\tau)$'s here are
\es{Yab3333}{
 {\bf 1} = [00]: \quad Y_{00}&=1\,,\\
 {\bf 10} = [02]: \quad Y_{10}&=\sigma-\tau\,,\\
 {\bf 14} = [20]: \quad Y_{11}&=\sigma+\tau-\frac25\,,\\
 {\bf 35'} = [04]: \quad Y_{20}&=\sigma^2+\tau^2-\frac23\sigma-\frac23\tau-2\sigma\tau+\frac16\,,\\
 {\bf 81} = [22]: \quad Y_{21}&=\sigma^2-\tau^2+\frac47\tau-\frac47\sigma\,,\\
 {\bf 55} = [40]: \quad Y_{22}&=\sigma^2+\tau^2- \frac{8}{9}\sigma  - \frac{8}{9} \tau + 
 4\sigma \tau +\frac{8}{63} \,,\\
 {\bf 84} = [06]: \quad Y_{30}&=\sigma^3 - \tau^3+\frac{2}{5}\sigma - \frac{6}{5}\sigma^2  - \frac{2}{5} \tau - 
 3\sigma^2 \tau + \frac{6}{5} \tau^2 + 3\sigma \tau^2\,,\\
 {\bf 220} = [24]: \quad Y_{31}&=\sigma^3 +\tau^3 +-\frac{4}{81} + \frac{34}{81}\sigma - \frac{32}{27}\sigma^2 + 
\frac{ 34}{81} \tau - \frac{8}{27}\sigma \tau -\sigma^2 \tau - 
\frac{ 32}{27} \tau^2 -\sigma \tau^2 \,,\\
 {\bf 260} = [42]: \quad Y_{32}&=\sigma^3-\tau^3+\frac{8}{33}\sigma - \frac{12}{11}\sigma^2  - \frac{8}{33} \tau + 
 3\sigma^2 \tau + \frac{12}{11} \tau^2 - 
 3\sigma \tau^2 \,,\\
 {\bf 140''} = [60]: \quad Y_{33}&=\sigma^3+\tau^3-\frac{16}{429} + \frac{72}{143}\sigma - \frac{18}{13}\sigma^2 \ + 
\frac{ 72}{143} \tau - \frac{72}{13}\sigma \tau + 9\sigma^2 \tau - \frac{18}{13} \tau^2 + 9\sigma \tau^2 \,.\\
}

If we now add up the two equations in \eqref{3D}, and expand in powers of $\bar\alpha$, then $z$ and $\bar z$ always appear in the combination $z^m+\bar{z}^m$ for some integer $m$, which can then be turned into rational functions of $U,V$. The resulting equation involves a set of differential operators in $U,V$ acting on ${\cal G}_{k,ab}(U,V)$, organized in powers of $\bar\alpha$. Finally, we convert the Ward identity to Mellin space by setting
\ie
{\cal G}_{k,ab }(U,V)\to M_{k,ab}(s,t),~~~U \partial_U \to \widehat{U\partial_U},~~~ V\partial_V \to \widehat{V\partial_V},~~~ U^mV^n \to \widehat{U^mV^n},
\fe
where the hatted operators act on $M _{ab}(s,t)$ as
\es{3DMellin}{
\widehat{U\partial_U} M _{k,ab}(s,t)&=\frac s 2  M _{k,ab}(s,t)\,,\\
\widehat{V\partial_V} M _{k,ab}(s,t)&=\left(\frac t 2-2k\right) M _{k,ab}(s,t)\,,\\
\widehat{U^mV^n} M _{k,ab}(s,t)&=M _{k,ab}(s-2m,t-2n)\left(2k-\frac{s}{2}\right)_m^2\left(2k-\frac{t}{2}\right)_n^2\left(2k-\frac{u}{2}\right)_{-m-n}^2\,,
}
where $u=8k-s-t$ and we will have independent constraints on each coefficient in the expansion in powers of $\bar\alpha$.

In position space, the Ward identities can be also be solved by writing $\mathcal{G}_k$ in the form \eqref{Ward2}, where the  differential operator $\Upsilon$ acts on $\mathcal{H}(U,V;\sigma,\tau)$ as
\es{thetaPos}{
\Upsilon&=\sigma^2\mathcal{D}'UV+\tau^2\mathcal{D}'U+\mathcal{D}'V-\sigma\mathcal{D}'V(U+1-V)
-\tau\mathcal{D}'(U+V-1)-\sigma\tau\mathcal{D}'U(V+1-U)\,,\\
\mathcal{D}'&=D-\frac{2}{V}(D^+_0-D^+_1+2\partial_\sigma \sigma)\tau\partial_\tau+\frac{2}{UV}(-V D_1^++2(V\partial_\sigma \sigma+\partial_{\tau}\tau-1))(\partial_\sigma\sigma+\partial_\tau\tau)\,,\\
D&=\partial_z\partial_{\bar z}-\frac{2}{z-\bar z}(\partial_z-\partial_{\bar z})\,,\\
D^+_0&=\partial_z+\partial_{\bar z}\,,\\
D^+_1&=z\partial_z+\bar z\partial_{\bar z}\,,
}
where $U=z\bar z$ and $V=(1-z)(1-\bar z)$.

The Mellin space version of this differential operator is a difference operator $\widehat{\Theta}$ that acts on 
\e{mtlmn}{\widetilde{M}_k(s,t;\sigma,\tau)\equiv\sum_{l+m+n=k-2}\sigma^m\tau^n\widetilde{M}^k_{lmn}(s,t)}
in the following way:
\es{theta6}{
&\widehat\Theta\circ\widetilde{M}^k_{lmn}(s,t)=-\frac{1}{4}\left(XY\widehat\Theta_4+XZ\widehat{V\Theta_4}+YZ\widehat{U\Theta_4}\right)\circ\widetilde{M}^k_{lmn}(s,t)\,,\\
&\Theta_4\equiv\tau+(1-\sigma-\tau)V+(-\tau-\sigma\tau+\tau^2)U+(\sigma^2-\sigma-\sigma\tau)UV+\sigma V^2+\sigma\tau U^2\,,\\
&X=s+4l-4k+2\,,\quad Y=t+4n-4k+2\,,\quad Z=u+4m-4k+2\,,
}
where $U,V$ acts on $\widetilde{M}_k(s,t)$ as
\es{3dMellin2}{
\widehat{U^mV^n} \circ\widetilde{M}_k(s,t)&=\widetilde{ M }_k(s-2m,t-2n)\left(2k-\frac{s}{2}\right)_m^2\left(2k-\frac{t}{2}\right)_n^2\left(2k-\frac{u}{2}\right)_{3-m-n}^2\,.
}
It is straightforward to take the flat space limit of this result directly; the result was given in \eqr{scwardflat}.

\sec{Superconformal blocks under chiral algebra twist}\label{appblocktwist}
In the chiral algebra limit \eqr{2dG}, the superconformal blocks for all multiplets in the $S_k\times S_k$ OPE \eqref{opemultEq} identically vanish except for the $\mathcal{D}[2n\, 0]$ and $\mathcal{B}[2n\, 0]$ multiplets, which take the simple form
\es{twistBlock}{
\mathfrak{G}_{\mathcal{D}[2n\,0]}(z\bar z,(1-z)(1-\bar z);{\bar z}^{-2},(1-{\bar z}^{-2})^2)=&\frac{4^n(\frac12)_n}{(1)_n}g_{4n,0}(z)\,,\\
\mathfrak{G}_{\mathcal{B}[2n\,0]_j}(z\bar z,(1-z)(1-\bar z);{\bar z}^{-2},(1-{\bar z}^{-2})^2)=&\frac{4^{n+1}(\frac12)_{n+1}}{(1)_{n+1}}A^{\mathcal{B}[2n\,0]}_{n+1\,n+1\,4n+6+j\,j+2}g_{4n+6+j,j+2}(z)\,,\\
}
where the $g_{\D,j}(z)$ are $SL(2,\mathbb{R})$ global conformal blocks 
\e{slt2B}{g_{\Delta,j}(z)=z^{\frac{\Delta+j}{2}}{}_2F_1\left(\frac{\Delta+j}{2},\frac{\Delta+j}{2},\Delta+j,z\right)\,,}
and $A^{\mathcal{B}[2n\,0]}_{n+1\,n+1\,4n+6+j\,j+2}$ is the coefficient defined in \eqref{GExpansion} that relates the superconformal descendent $(4n+6+j,j+2)_{[2n\,0]}$ to the superconformal primary $(4n+4+j,j)_{[2n\,0]}$ of $\mathcal{B}[2n\, 0]_j$.\footnote{As shown in \cite{Buican:2016hpb}, such a superconformal descendant always exists. For $k=2$, for instance, we list the conformal primaries that appear in $\mathcal{B}[20]_j$ explicitly in Appendix \ref{super20}.} The prefactors in \eqref{twistBlock} come from the $R$-symmetry factors $Y_{ab}(\sigma,\tau)$ that multiply the conformal block $G_{\Delta',j'}(U,V)$ of each superconformal descendent that appears in the superconformal block $\mathfrak{G}_{\Delta,j}(U,V;\sigma,\tau)$ in \eqref{GExpansion}. After performing the twist and expanding for small $\bar z$, these quantities take the form
\es{2dto6d}{
\lim_{\bar z\to0}Y_{nn}(\bar z^{-2},(1-\bar z^{-2})^2)&= \bar z^{-2n}\left[ \frac{4^n(1/2)_n}{(1)_n}+O(\bar z^{-1})\right]\,,\\
\lim_{\bar z\to0} G_{\Delta,j}(z \bar z,(1-z)(1-\bar z))& = \bar z^{\frac{\Delta-j}{2}}\left[g_{\Delta,j}(z)+O(\bar z^{-1})\right]\,.
}
Since the $\bar z$ dependence must cancel from the superblock after performing the twist, we see that for $\mathcal{D}[2n\, 0]$ only the superconformal primary with $Y_{nn}(\sigma,\tau)G_{4n,0}(U,V)$ survives, while for $\mathcal{B}[2n\, 0]_j$ only the superconformal descendent  with $Y_{n+1\,n+1}(\sigma,\tau)G_{4n+6+j,j+2}(U,V)$ survives. Putting things together, $\Gc_k(z)|_{2d}$ takes the form \eqr{2dblockExp} given in the main text.

\section{Mack polynomials}
\label{Mack}

We use the definition of the Mack polynomials in \cite{Gopakumar:2016cpb}, but with a different convention for $s$ and $t$, namely $s_\text{there}=s_\text{here}/2$ and $t_\text{there}=t_\text{here}/2-(\Delta_2+\Delta_3)/2$, and we (anti)-symmetrize $t,u$ for (odd) even $j$. For a four-point function of identical scalars with dimension $\Delta$ we then define
\es{Qdefine}{
&\mathcal{Q}_{j',m}^{\Delta',\Delta}(t)=\mathfrak{N}_{\Delta,j,d}\frac{(-1)^m4^{j'}\sin\pi(\Delta'-d/2)\Gamma(d/2-\Delta'-m)}{m!(\Delta'-1)_{j'}(d-\Delta'-1)_{j'}\Gamma(\frac{2\Delta-\Delta'+j'-2m}{2})^2}\\
&\times \frac{P_{\Delta'-d/2,j'}((\Delta'-j')/2+m,t/2-\Delta)+(-1)^j P_{\Delta'-d/2,j'}((\Delta'-j')/2+m,u/2-\Delta)}{2}\,,
}
where $d$ is the spacetime dimension of the CFT, $\mathfrak{N}_{\Delta,j,d}$ is a normalization factor, and $P_{\nu,j}(s,t)$ is
\es{Pdefine}{
&P_{\nu,j}(s,t)=\widetilde{\sum}\frac{\Gamma(\lambda_1)^2\Gamma(\bar\lambda_1)^2(\lambda_2-s)_k(\bar\lambda_2-s)_k(s+t)_\beta(s+t)_\alpha(-t)_{m-\alpha}(-t)_{j-2k-m-\beta}  }{\prod_i\Gamma(l_i)}\,,\\
&\widetilde{\sum}\equiv\frac{j!}{2^j(d/2-1)_j}\sum_{k=0}^{\left\lfloor{ j/2}\right\rfloor}\sum_{m=0}^{j-2k}\sum_{\alpha=0}^m\sum_{\beta=0}^{j-2k-m}\frac{(-1)^{j-k-\alpha-\beta}\Gamma(j-k+d/2-1)}{\Gamma(d/2-1)k!(j-2k)!}\begin{pmatrix}j-2k\\m\end{pmatrix}\begin{pmatrix}m\\ \alpha\end{pmatrix}\begin{pmatrix}j-2k-m\\\beta\end{pmatrix}\,,\\
&\lambda_1=\frac{d/2+\nu+j}{2}\,,\quad \bar\lambda_1=\frac{d/2-\nu+j}{2}\,,\quad \lambda_2=\frac{d/2+\nu-j}{2}\,,\quad \bar\lambda_2=\frac{d/2-\nu-j}{2}\,,\\
&l_1=\lambda_2+j-k-m+\alpha-\beta\,,\quad l_2=\lambda_2+k+m-\alpha+\beta\,,\quad l_3=\bar\lambda_2+k+m\,,\quad l_4=\bar\lambda_2+j-k-m\,.
}
The maximal value $m_\text{max}=\Delta-(\Delta'-j')/2-1$ in the sum over $m$ in \eqref{exchange} is here explicitly enforced by the Gamma function in the numerator of \eqref{Qdefine}. Since $\sum_{m=0}^{m_\text{max}}\frac{\mathcal{Q}_{j',m}^{\Delta',\Delta}(t)}{s-(\Delta'-j')}$ is defined to have the same meromorphic part as the Mellin transform of a conformal block $G_{\Delta',j'}(U,V)$, we fix the normalization factor $\mathfrak{N}_{\Delta,j,d}$ so that in the limit $U\to0,V\to1$ we recover our normalization of the conformal blocks $U^\frac{\Delta'-j'}{2}(1-V)^{j'}$. We can fix this by taking the residue of the pole $s=\Delta'-j'$ and then summing over the $t$ poles in the inverse Mellin transform \eqref{mellinDef} of $\sum_{m=0}^{m_\text{max}}\frac{\mathcal{Q}_{j',m}^{\Delta',\Delta}(t)}{s-(\Delta'-j')}$ and expanding around $V=1$. For the blocks considered in this paper, we find
\es{normR}{
\mathfrak{N}_{4,0}&=\frac{12}{\pi}\,,\quad \mathfrak{N}_{5,1}=\frac{60}{\pi}\,,\quad \mathfrak{N}_{6,2}=\frac{350}{\pi}\,,\\ \mathfrak{N}_{8,0}&=\frac{2800}{3\pi}\,,\quad \mathfrak{N}_{9,1}=\frac{6300}{\pi}\,,\quad \mathfrak{N}_{10,0}=\frac{11025}{\pi}\,,\quad \mathfrak{N}_{10,2}=\frac{43659}{\pi}\,,\quad \mathfrak{N}_{11,1}=\frac{77616}{\pi}\,.
}

\section{Polynomial Mellin amplitudes}
\label{amps}

We now give explicit expressions for various Mellin amplitudes used in the main text

The polynomial parts $M_{k,\text{poly}}^{(1)}$ of the supergravity meromorphic amplitudes $M_{k,\,\rm sugra}\equiv M_{k,\rm mero}^{(1)}$ in Section \ref{tree} are 
\es{polyParts}{
M_{2,\,\text{poly}}^{(1)}=&\frac{1}{8}  (5 (16- t) \sigma^2 - 
   5 s ( \sigma-1) (1 + \sigma - 4\tau) + (96 + 
      5 t (\tau-4)) \tau + 4 \sigma (5 t - 56 (1 +\tau)))\,,\\
M_{3,\,\text{poly}}^{(1)}=&-\frac{1}{72}  \left(3 s (\sigma -1) \left(\sigma ^2+\sigma  (2-3 \tau )-4 \tau ^2-3 \tau
   +1\right)+3 \sigma ^3 (t-24)\right.\\
& \left.  +\sigma ^2 (3 t (\tau -3)+124 (\tau +1))+\sigma  \left(4 \left(49 \tau
   ^2+136 \tau +49\right)-3 t \left(\tau ^2+4\right)\right)\right.\\
& \left.  -\tau  (\tau +1) (3 t (\tau -4)+92)\right)\,,
   }
The polynomial part $M_{3,\text{poly}}^{(4)}\subset M_{3,\rm mero}^{(4)}$ is
   \es{polyPart2}{
M_{3,\text{poly}}^{(4)}=&\frac{1}{165} \left(165 s^4 \tau ^2 (\sigma +\tau +1)+55 s^3 \tau  \left(-5 \sigma ^2-145 \sigma  \tau
   -144 \tau ^2+6 t \left((\tau +1)^2-\sigma ^2\right)-143 \tau +5\right) \right.\\
 & \left. +5 s^2 \left(4092 \sigma ^3+6
   \sigma ^2 (365 \tau -682)+2 \sigma  \left(12297 \tau ^2-8306 \tau -2046\right) \right.\right.\\
 &\left.  \left. +33 t^2 \left(\sigma
   ^3-\sigma ^2 (\tau +1)-\sigma  \left(\tau ^2+1\right)+\tau ^3+5 \tau ^2+5 \tau +1\right)+6
   \left(4732 \tau ^3+3967 \tau ^2-295 \tau +682\right)\right.\right.\\
&\left.\left.   -11 t \left(67 \sigma ^3-\sigma ^2 (207 \tau
   +67)-\sigma  \left(71 \tau ^2+285 \tau +67\right)+211 \tau ^3+358 \tau ^2+210 \tau
   +67\right)\right) \right. \\
 & \left. +48 \left(53316 \sigma ^3+11417 \sigma ^2 (\tau +1)+7 \sigma  \left(2621 \tau
   ^2-23380 \tau +2621\right)+69480 \tau ^3-5161 \tau ^2-5161 \tau \right.\right.\\
 & \left.\left. +69480\right)+s \left(-12
   \left(38226 \sigma ^3+3 \sigma ^2 (3821 \tau -14025)+\sigma  \left(57138 \tau ^2-166120 \tau
   -39765\right)  \right.\right.\right.\\
&\left.\left.\left.   +93840 \tau ^3+46662 \tau ^2-7263 \tau +43614\right)-330 t^3 \left(\sigma ^2-(\tau
   +1)^2\right)-55 t^2 \left(67 \sigma ^3-\sigma ^2 (67 \tau +207)\right. \right. \right.\\
 &\left.\left. \left. -\sigma  \left(67 \tau ^2+285 \tau
   +71\right)+67 \tau ^3+210 \tau ^2+358 \tau +211\right)+20 t \left(4122 \sigma ^3-6558 \sigma ^2
   (\tau +1) \right. \right.\right.\\
 & \left. \left.\left.-\sigma  \left(4332 \tau ^2+22963 \tau +4332\right)+6768 \tau ^3+6861 \tau ^2+6861 \tau
   +6768\right)\right)+165 t^4 (\sigma +\tau +1)\right.\\
 &\left.  -55 t^3 \left(5 \sigma ^2+145 \sigma -5 \tau ^2+143
   \tau +144\right)+10 t^2 \left(2046 \sigma ^3-3 \sigma ^2 (682 \tau -365) \right.\right.\\
&\left. \left.  +\sigma  \left(-2046 \tau
   ^2-8306 \tau +12297\right)+3 \left(682 \tau ^3-295 \tau ^2+3967 \tau +4732\right)\right) \right.\\
&\left.   -12 t
   \left(38226 \sigma ^3+\sigma ^2 (11463-42075 \tau )+\sigma  \left(-39765 \tau ^2-166120 \tau
   +57138\right)+43614 \tau ^3 \right.\right.\\
&\left.\left.   -7263 \tau ^2+46662 \tau +93840\right)\right)\,.\\
}
Finally, we write the reduced Mellin amplitude in the notation of \eqref{3333Sugra} for the lowest few polynomial amplitudes for $k=3$:
   \es{polyPart3}{
   \widetilde{M}_{100}^{(5,1)}(s,t)=&\frac{128}{13 ( t-10) ( 8-s - t))}\,,\\
     \widetilde{M}_{100}^{(6,1)}(s,t)=&-(128 (-8 (-10 + s) (-220707864 + 
          s (85593762 + s (-9978487 + s (250829 + 11193 s)))) \\
 &         + (-18 +
           s) (-154429232 + 
          s (63343800 + 
             s (-7510574 + 3 s (53379 + 3731 s)))) t \\
 &            + (113748160 + 
          3 s (-9833184 + s (1316254 + s (-145799 + 7462 s)))) t^2 \\
&          +        574 (-18 + s) (2884 - 678 s + 39 s^2) t^3 + 
       287 (2884 - 678 s + 39 s^2) t^4) )\\
 &    \times  (78351 (-10 + s) (-8 + s) (-6 + s) (-10 + t) (-8 + t) (-10 +
        s + t) (-8 + s + t))^{-1}\,,\\
  \widetilde{M}_{100}^{(6,2)}(s,t)=&-(64 (-8 (-10 + s) (-30379752 + 
          s (12544806 + s (-1668613 + 71813 s))) \\
          &+ (-18 + 
          s) (-20911376 + 
          s (9366240 + s (-1316054 + 58857 s))) t \\
 &         + (17399680 + 
          3 s (-1299032 + s (106998 + s (-8835 + 533 s)))) t^2\\
&           + 
       82 (-18 + s) (2884 - 678 s + 39 s^2) t^3 + 
       41 (2884 - 678 s + 39 s^2) t^4) )\\
&       (11193 (-10 + s) (-8 + s) (-6 + s) (-10 + t) (-8 + t) (-10 +
        s + t) (-8 + s + t))^{-1}\,.
   }
   
   \section{Supermultiplets and superblocks in $S_2\times S_2$}
\label{super20}

In this appendix we discuss the supermultiplets that appear in $S_2\times S_2$. First, we list the conformal primaries that appear in each supermultiplet. Following the algorithm in \cite{Buican:2016hpb}, we list these results in Tables \ref{D20}-\ref{A00}. 

\begin{table}[htpb]
\centering
\begin{tabular}{|c||c|c|c|c|c|c|}
\hline 
$\mathcal{D}{[20]}$ & \multicolumn{6}{c|}{spins in various $\mathfrak{so}(5)$ irreps}  \\
\hline
\multirow{2}{*}{dimension}  & ${\bf 1}$ & ${\bf 10}$ & ${\bf 14}$ & ${\bf 35'}$ & ${\bf 81}$ & ${\bf 55}$\\  
 & $[00]$ & $[02]$ & $[20]$ & $[04]$ & $[22]$ & $[40]$\\          
\hline
$4$       & --          & --          & 0          & --          & --          & --    \\
$5$       & --          & 1          & --          & --          & --         & --     \\
$6$       & 2          & --          & --         & --         & --          & --     \\
\hline
\end{tabular}
\caption{All possible conformal primaries in $\mathcal{D}{[20]}\times \mathcal{D}{[20]}$ corresponding to the $\mathcal{D}{[20]}$ superconformal multiplet.}\label{D20}
\end{table}

\begin{table}[htpb]
\centering
\begin{tabular}{|c||c|c|c|c|c|c|}
\hline 
$\mathcal{D}{[40]}$ & \multicolumn{6}{c|}{spins in various $\mathfrak{so}(5)$ irreps}  \\
\hline
\multirow{2}{*}{dimension}  & ${\bf 1}$ & ${\bf 10}$ & ${\bf 14}$ & ${\bf 35'}$ & ${\bf 81}$ & ${\bf 55}$\\  
 & $[00]$ & $[02]$ & $[20]$ & $[04]$ & $[22]$ & $[40]$\\          
\hline
$8$       & --          & --          & --          & --          & --          & 0    \\
$9$       & --          & --          & --          & --          & 1         & --     \\
$10$       & --          & --          & 2         & 0         & --          & --     \\
$11$       & --          & 1          & --         & --         & --          & --     \\
$12$       & 0          & --          & --         & --         & --          & --     \\
\hline
\end{tabular}
\caption{All possible conformal primaries in $\mathcal{D}{[20]}\times \mathcal{D}{[20]}$ corresponding to the $\mathcal{D}{[40]}$ superconformal multiplet.}\label{D40}
\end{table}

\begin{table}[htpb]
\centering
\begin{tabular}{|c||c|c|c|c|c|c|}
\hline 
$\mathcal{D}{[04]}$ & \multicolumn{6}{c|}{spins in various $\mathfrak{so}(5)$ irreps}  \\
\hline
\multirow{2}{*}{dimension}  & ${\bf 1}$ & ${\bf 10}$ & ${\bf 14}$ & ${\bf 35'}$ & ${\bf 81}$ & ${\bf 55}$\\  
 & $[00]$ & $[02]$ & $[20]$ & $[04]$ & $[22]$ & $[40]$\\          
\hline
$8$       & --          & --          & --          & 0          & --          & --    \\
$9$       & --          & 1          & --          & --          & 1         & --     \\
$10$       & 0          & --          & 2,0         & 2         & --          & 0     \\
$11$       & --          & 1,3          & --         & --         & 1          & --     \\
$12$       & 2          & --          & 2         & 0         & --          & --     \\
$13$       & --          & 1          & --         & --         & --          & --     \\
$14$       & 0          & --          & --         & --         & --          & --     \\
\hline
\end{tabular}
\caption{All possible conformal primaries in $\mathcal{D}{[20]}\times \mathcal{D}{[20]}$ corresponding to the $\mathcal{D}{[04]}$ superconformal multiplet.}\label{D04}
\end{table}

\begin{table}[htpb]
\centering
\begin{tabular}{|c||c|c|c|c|c|c|}
\hline 
$\mathcal{B}{[02]}_j$ & \multicolumn{6}{c|}{spins in various $\mathfrak{so}(5)$ irreps}  \\
\hline
\multirow{2}{*}{dimension}  & ${\bf 1}$ & ${\bf 10}$ & ${\bf 14}$ & ${\bf 35'}$ & ${\bf 81}$ & ${\bf 55}$\\  
 & $[00]$ & $[02]$ & $[20]$ & $[04]$ & $[22]$ & $[40]$\\          
\hline
$j+8$       & --          & $j$          & --          & --          & --          & --    \\
$j+9$       & $j\pm1$          & --          & $j\pm1$          & $j+1$          & --         & --     \\
$j+10$       & --          & $j,j\pm2$          & --         & --         & $j,j+2$          & --     \\
$j+11$       & $j\pm1,j\pm3$          & --          & $j\pm1,j+3$         & $j\pm1,j+3$         & --          & $j+1$     \\
$j+12$       & --          & $j,j\pm2,j+4$          & --         & --         & $j,j+2$          & --     \\
$j+13$       & $j\pm1.j+3$          & --          & $j\pm1,j+3$         & $j+1$         & --          & --     \\
$j+14$       & --          & $j,j+2$          & --         & --         & --          & --     \\
$j+15$       & $j+1$          & --         & --         & --         & --          & --     \\
\hline
\end{tabular}
\caption{All possible conformal primaries in $\mathcal{D}{[20]}\times \mathcal{D}{[20]}$ corresponding to the $\mathcal{B}{[02]}_j$ superconformal multiplet.}\label{B02}
\end{table}

\begin{table}[htpb]
\centering
\begin{tabular}{|c||c|c|c|c|c|c|}
\hline 
$\mathcal{B}{[20]}_j$ & \multicolumn{6}{c|}{spins in various $\mathfrak{so}(5)$ irreps}  \\
\hline
\multirow{2}{*}{dimension}  & ${\bf 1}$ & ${\bf 10}$ & ${\bf 14}$ & ${\bf 35'}$ & ${\bf 81}$ & ${\bf 55}$\\  
 & $[00]$ & $[02]$ & $[20]$ & $[04]$ & $[22]$ & $[40]$\\          
\hline
$j+8$       & --               & --              & $j$                & --          & --          & --    \\
$j+9$       & --               & $j\pm1$    & --                  & --          & $j+1$         & --     \\
$j+10$       & $j,j\pm2$ & --               & $j,j+2$         & $j,j+2$ & --          & $j+2$     \\
$j+11$       & --               & $j\pm1,j+3$  & --              & --         & $j+1,j+3$          & --     \\
$j+12$       & $j,j+2$      & --                & $j,j+2,j+4$   & $j+2$  & --          & --     \\
$j+13$       & --               & $j+1,j+3$    & --                 & --        & --          & --     \\
$j+14$       & $j+2$         & --                & --                 & --         & --          & --     \\
\hline
\end{tabular}
\caption{All possible conformal primaries in $\mathcal{D}{[20]}\times \mathcal{D}{[20]}$ corresponding to the $\mathcal{B}{[20]}_j$ superconformal multiplet.}\label{B20}
\end{table}

\begin{table}[htpb]
\centering
\begin{tabular}{|c||c|c|c|c|c|c|}
\hline 
$\mathcal{A}{[00]}_{\Delta,j}$ & \multicolumn{6}{c|}{spins in various $\mathfrak{so}(5)$ irreps}  \\
\hline
\multirow{2}{*}{dimension}  & ${\bf 1}$ & ${\bf 10}$ & ${\bf 14}$ & ${\bf 35'}$ & ${\bf 81}$ & ${\bf 55}$\\  
 & $[00]$ & $[02]$ & $[20]$ & $[04]$ & $[22]$ & $[40]$\\          
\hline
$\Delta$           & $j$                        & --                      & --               & --              & --              & --    \\
$\Delta+1$       & --                          & $j\pm1$             & --               & --             & --              & --    \\
$\Delta+2$       & $j,j\pm2$               & --                      & $j,j\pm2$  & $j$            & --   & --    \\
$\Delta+3$       & --                          & $j\pm1,j\pm3$  & --               & --               & $j\pm1$              & --    \\
$\Delta+4$       & $j,j\pm2,j\pm4$    & --                       & $j,j\pm2$  & $j,j\pm2$   & --   & $j$    \\
$\Delta+5$       & --                          & $j\pm1,j\pm3$   & --              & --                & $j\pm1$             & \\
$\Delta+6$       & $j,j\pm2$             & --                       & $j,j\pm2$   & $j$             & --             & --    \\
$\Delta+7$       & --                          & $j\pm1$            & --               & --                & --            &   \\
$\Delta+8$       & $j$                        & --                       & --              & --               & --             &  \\
\hline
\end{tabular}
\caption{All possible conformal primaries in $\mathcal{D}{[20]}\times \mathcal{D}{[20]}$ corresponding to the $\mathcal{A}{[00]}_{\Delta,j}$ superconformal multiplet.}\label{A00}
\end{table}

Note that the superconformal primary for $\mathcal{B}[02]_j$ also appear as a superconformal descendent in $\mathcal{A}[00]_{\Delta,j}$, but that the $(j+11,j+1)_{[40]}$ conformal primary does not appear in $\mathcal{A}[00]_{\Delta,j}$ or $\mathcal{D}[04]$ for any $j$ or $\Delta\geq j+8$. The coefficient $A^{\mathcal{B}[02]_j}_{10\,j+11\,j+1}$ that relates this conformal primary to the superconformal primary can be computed by plugging a linear combination of conformal blocks for each conformal primary appearing in these tables into the superconformal Ward identities \eqref{3D}, which fixes all such relative coefficients. We define the conformal blocks with an extra factor of $(-2)^j$ relative to \cite{Beem:2015aoa}:
\es{conformalBlock}{
G_{\Delta,j}(U,V)=&\mathcal{F}_{00}-\frac{j+3}{j+1}\mathcal{F}_{-11}+\frac{(\Delta-4)(j+3)(\Delta-j-4)^2}{16(\Delta-2)(j+1)(\Delta-j-5)(\Delta-j-3)}\mathcal{F}_{02}\\
&-\frac{(\Delta-4)(\Delta+j)^2}{16(\Delta-2)((\Delta+j)^2-1)}\mathcal{F}_{11}\,,\\
\mathcal{F}_{nm}(z,\bar z)\equiv &\frac{(z \bar z)^{\frac{\Delta-j}{2}}}{(z-\bar z)^3}\left({z}^jz^{n+3}\bar z^m{}_2F_1\left(\frac{\Delta+j}{2}+n,\frac{\Delta+j}{2}+n,\Delta+j+2n,z\right)\right.\\
&\left. {}_2F_1\left(\frac{\Delta-j}{2}-3+m,\frac{\Delta-j}{2}-3+m,\Delta-j-6+2m,\bar z\right)-(z\leftrightarrow \bar z)\right)\,,
}
where recall that $U=z \bar z$ and $V=(1-z)(1-\bar z)$. If we normalize the superconformal primary to have unit coefficient, then we find that $A^{\mathcal{B}[02]_j}_{10\,j+11\,j+1}$ for $j=1,3,5$ are
\es{coeffs}{
A^{\mathcal{B}[02]_1}_{10\,12\,2}= -\frac{3}{5}\,,\qquad A^{\mathcal{B}[02]_3}_{10\,14\,4}= -\frac{10}{21} \,,\qquad A^{\mathcal{B}[02]_5}_{10\,16\,6} = -\frac{35}{81}\,.  \\
}

\sec{$R^4$ and $D^6R^4$ coefficients in M-theory}\label{appstring}

Here we perform the derivation of the finite contributions of $R^4$ and $D^6R^4$ to the 11d four-point superamplitude. We eschew the 11d action and instead work with type IIA amplitudes, which we uplift to 11d. The following relations are useful:
\e{relns}{e^{2\phi}(\a')^3 = {\ell_{11}^6\o (2\pi)^2}~, \quad \left({L_{\rm AdS}\o \ell_{11}}\right)^9 \approx 16c\,,}
where $e^\phi = g_s$. We denote IIA amplitudes as $A$ and 11d amplitudes as $\Ac^{11}$.

From the tree-level amplitude of type IIA \c{Green:1987sp},
\e{atreeIIA}{A_{\tree} = \widehat K \kappa_{10}^2 e^{-2\phi} {2^6\o (\a')^3 stu}\exp\left[\sum_{k=1}^\i {2\zeta(2k+1)\o 2k+1}(\a'/4)^{2k+1}(s^{2k+1}+t^{2k+1}+u^{2k+1})\right]\,,}
where $u=-s-t$. We have, up to the same universal coefficient $\widehat K\kappa_{10}^2$,
\es{}{A_{\tree}\Big|_{R} &= {64\o (\a')^3 stu}e^{-2\phi}\,,\\
A_{\tree}\Big|_{R^4} &= 2\zeta(3)e^{-2\phi}\,,\\
A_{\tree}\Big|_{D^6R^4} &= {(\a')^3 stu\o 32}\zeta(3)^2e^{-2\phi}\,.}

First we address $R^4$. From the type IIA action \c{Green:1997tv},
\e{}{S_{R^4} \propto 2\zeta(3) E_{3/2}(\phi) = 2\zeta(3) e^{-3\phi/2}(1+{\pi^2\o 3\zeta(3)} e^{2\phi}) + (\text{non-perturbative})\,,}
which implies
\e{}{{A_{\rm 1-loop}|_{R^4} \o A_{\tree}|_{R} } = g_s^2{\pi^2\o 3\zeta(3)}{A_{\tree}|_{R^4} \o A_{\tree}|_{R} } =g_s^2 (\a')^3 {\pi^2\o 96}stu\,.}
This term is finite in the uplift to 11d, as it is independent of $R_{11}$. Uplifting to 11d using \eqr{relns},
\e{}{{\Ac^{11}|_{R^4} \o \Ac^{11}|_{R} } = \ell_{11}^6{stu\o 3\cdot 2^7}\,.}
Next, we have \c{Green:2005ba}
\e{}{S_{D^6R^4} \propto 4\zeta(3)^2e^{-2\phi} + 8\zeta(2)\zeta(3)+{48\o 5}\zeta(2)^2e^{2\phi} + {8\o 9}\zeta(6) e^{4\phi} + (\text{non-perturbative})\,,}
where $\zeta(2) = \pi^2/90$ and $\zeta(6)=\pi^6/945$. The two-loop term gives rise to a finite term in 11d,
\e{}{{A_{\rm 2-loop}|_{D^6R^4} \o A_{\tree}|_{R} } = g_s^4{12 \zeta(2)^2 \o 5 \zeta(3)^2}{A_{\tree}|_{D^6R^4} \o A_{\tree}|_{R} } =g_s^4 (\a')^6 \,{3\zeta(2)^2\o 2560}(stu)^2\,.}
Uplifting to 11d using \eqr{relns},
\e{}{{\Ac^{11}|_{D^6R^4} \o \Ac^{11}|_{R} } = \ell_{11}^{12}{(stu)^2\o 15\cdot 2^{15}}\,.}
Notice that this depends only on $(stu)^2$, not $(s^2+t^2+u^2)^3$. 

\bibliographystyle{ssg}
\bibliography{6dDraftbib}

\begingroup\raggedright\begin{thebibliography}{10}

\bibitem{Beem:2014kka}
C.~Beem, L.~Rastelli, and B.~C. van Rees, ``{$ \mathcal{W} $ symmetry in six
  dimensions},'' {\em JHEP} {\bf 05} (2015) 017,
  \href{http://xxx.lanl.gov/abs/1404.1079}{{\tt 1404.1079}}.

\bibitem{Witten:1995zh}
E.~Witten, ``{Some comments on string dynamics},'' in {\em {Future perspectives
  in string theory. Proceedings, Conference, Strings'95, Los Angeles, USA,
  March 13-18, 1995}}, pp.~501--523, 1995.
\newblock \href{http://xxx.lanl.gov/abs/hep-th/9507121}{{\tt hep-th/9507121}}.

\bibitem{Seiberg:1997ax}
N.~Seiberg, ``{Notes on theories with 16 supercharges},'' {\em Nucl. Phys.
  Proc. Suppl.} {\bf 67} (1998) 158--171,
  \href{http://xxx.lanl.gov/abs/hep-th/9705117}{{\tt hep-th/9705117}}.

\bibitem{Aharony:1997th}
O.~Aharony, M.~Berkooz, S.~Kachru, N.~Seiberg, and E.~Silverstein, ``{Matrix
  description of interacting theories in six-dimensions},'' {\em Adv. Theor.
  Math. Phys.} {\bf 1} (1998) 148--157,
  \href{http://xxx.lanl.gov/abs/hep-th/9707079}{{\tt hep-th/9707079}}.

\bibitem{Lambert:2010iw}
N.~Lambert, C.~Papageorgakis, and M.~Schmidt-Sommerfeld, ``{M5-Branes,
  D4-Branes and Quantum 5D super-Yang-Mills},'' {\em JHEP} {\bf 01} (2011) 083,
  \href{http://xxx.lanl.gov/abs/1012.2882}{{\tt 1012.2882}}.

\bibitem{Douglas:2010iu}
M.~R. Douglas, ``{On D=5 super Yang-Mills theory and (2,0) theory},'' {\em
  JHEP} {\bf 02} (2011) 011, \href{http://xxx.lanl.gov/abs/1012.2880}{{\tt
  1012.2880}}.

\bibitem{Maldacena:1997re}
J.~M. Maldacena, ``{The Large N limit of superconformal field theories and
  supergravity},'' {\em Int. J. Theor. Phys.} {\bf 38} (1999) 1113--1133,
  \href{http://xxx.lanl.gov/abs/hep-th/9711200}{{\tt hep-th/9711200}}. [Adv.
  Theor. Math. Phys.2,231(1998)].

\bibitem{Gubser:1998bc}
S.~S. Gubser, I.~R. Klebanov, and A.~M. Polyakov, ``{Gauge theory correlators
  from noncritical string theory},'' {\em Phys. Lett.} {\bf B428} (1998)
  105--114, \href{http://xxx.lanl.gov/abs/hep-th/9802109}{{\tt
  hep-th/9802109}}.

\bibitem{Witten:1998qj}
E.~Witten, ``{Anti-de Sitter space and holography},'' {\em Adv. Theor. Math.
  Phys.} {\bf 2} (1998) 253--291,
  \href{http://xxx.lanl.gov/abs/hep-th/9802150}{{\tt hep-th/9802150}}.

\bibitem{Beem:2015aoa}
C.~Beem, M.~Lemos, L.~Rastelli, and B.~C. van Rees, ``{The (2, 0)
  superconformal bootstrap},'' {\em Phys. Rev.} {\bf D93} (2016), no.~2 025016,
  \href{http://xxx.lanl.gov/abs/1507.05637}{{\tt 1507.05637}}.

\bibitem{Pilch:1984xy}
K.~Pilch, P.~van Nieuwenhuizen, and P.~K. Townsend, ``{Compactification of
  $d=11$ Supergravity on S(4) (Or 11 = 7 + 4, Too)},'' {\em Nucl. Phys.} {\bf
  B242} (1984) 377--392.

\bibitem{vanNieuwenhuizen:1984iz}
P.~van Nieuwenhuizen, ``{The Complete Mass Spectrum of $d=11$ Supergravity
  Compactified on S(4) and a General Mass Formula for Arbitrary Cosets M(4)},''
  {\em Class. Quant. Grav.} {\bf 2} (1985) 1.

\bibitem{Nastase:1999cb}
H.~Nastase, D.~Vaman, and P.~van Nieuwenhuizen, ``{Consistent nonlinear K K
  reduction of 11-d supergravity on AdS(7) x S(4) and selfduality in odd
  dimensions},'' {\em Phys. Lett.} {\bf B469} (1999) 96--102,
  \href{http://xxx.lanl.gov/abs/hep-th/9905075}{{\tt hep-th/9905075}}.

\bibitem{Nastase:1999kf}
H.~Nastase, D.~Vaman, and P.~van Nieuwenhuizen, ``{Consistency of the AdS(7) x
  S(4) reduction and the origin of selfduality in odd dimensions},'' {\em Nucl.
  Phys.} {\bf B581} (2000) 179--239,
  \href{http://xxx.lanl.gov/abs/hep-th/9911238}{{\tt hep-th/9911238}}.

\bibitem{Prochazka:2014gqa}
T.~Procházka, ``{Exploring $ {\mathcal{W}}_{\infty } $ in the quadratic
  basis},'' {\em JHEP} {\bf 09} (2015) 116,
  \href{http://xxx.lanl.gov/abs/1411.7697}{{\tt 1411.7697}}.

\bibitem{Gaberdiel:2012ku}
M.~R. Gaberdiel and R.~Gopakumar, ``{Triality in Minimal Model Holography},''
  {\em JHEP} {\bf 07} (2012) 127, \href{http://xxx.lanl.gov/abs/1205.2472}{{\tt
  1205.2472}}.

\bibitem{Beem:2013sza}
C.~Beem, M.~Lemos, P.~Liendo, W.~Peelaers, L.~Rastelli, and B.~C. van Rees,
  ``{Infinite Chiral Symmetry in Four Dimensions},'' {\em Commun. Math. Phys.}
  {\bf 336} (2015), no.~3 1359--1433,
  \href{http://xxx.lanl.gov/abs/1312.5344}{{\tt 1312.5344}}.

\bibitem{Chester:2014mea}
S.~M. Chester, J.~Lee, S.~S. Pufu, and R.~Yacoby, ``{Exact Correlators of BPS
  Operators from the 3d Superconformal Bootstrap},'' {\em JHEP} {\bf 03} (2015)
  130, \href{http://xxx.lanl.gov/abs/1412.0334}{{\tt 1412.0334}}.

\bibitem{Dedushenko:2016jxl}
M.~Dedushenko, S.~S. Pufu, and R.~Yacoby, ``{A one-dimensional theory for Higgs
  branch operators},'' \href{http://xxx.lanl.gov/abs/1610.00740}{{\tt
  1610.00740}}.

\bibitem{Bastianelli:1999en}
F.~Bastianelli and R.~Zucchini, ``{Three point functions of chiral primary
  operators in d = 3, N=8 and d = 6, N=(2,0) SCFT at large N},'' {\em Phys.
  Lett.} {\bf B467} (1999) 61--66,
  \href{http://xxx.lanl.gov/abs/hep-th/9907047}{{\tt hep-th/9907047}}.

\bibitem{Campoleoni:2011hg}
A.~Campoleoni, S.~Fredenhagen, and S.~Pfenninger, ``{Asymptotic W-symmetries in
  three-dimensional higher-spin gauge theories},'' {\em JHEP} {\bf 09} (2011)
  113, \href{http://xxx.lanl.gov/abs/1107.0290}{{\tt 1107.0290}}.

\bibitem{Cordova:2016cmu}
C.~Cordova and D.~L. Jafferis, ``{Toda Theory From Six Dimensions},'' {\em
  JHEP} {\bf 12} (2017) 106, \href{http://xxx.lanl.gov/abs/1605.03997}{{\tt
  1605.03997}}.

\bibitem{Grisaru:1976vm}
M.~T. Grisaru, H.~N. Pendleton, and P.~van Nieuwenhuizen, ``{Supergravity and
  the S Matrix},'' {\em Phys. Rev.} {\bf D15} (1977) 996.

\bibitem{Tseytlin:2000sf}
A.~A. Tseytlin, ``{R**4 terms in 11 dimensions and conformal anomaly of (2,0)
  theory},'' {\em Nucl. Phys.} {\bf B584} (2000) 233--250,
  \href{http://xxx.lanl.gov/abs/hep-th/0005072}{{\tt hep-th/0005072}}.

\bibitem{Green:1997as}
M.~B. Green, M.~Gutperle, and P.~Vanhove, ``{One loop in eleven-dimensions},''
  {\em Phys. Lett.} {\bf B409} (1997) 177--184,
  \href{http://xxx.lanl.gov/abs/hep-th/9706175}{{\tt hep-th/9706175}}.
  [,164(1997)].

\bibitem{Green:1999pu}
M.~B. Green, H.-h. Kwon, and P.~Vanhove, ``{Two loops in eleven-dimensions},''
  {\em Phys. Rev.} {\bf D61} (2000) 104010,
  \href{http://xxx.lanl.gov/abs/hep-th/9910055}{{\tt hep-th/9910055}}.

\bibitem{Russo:1997mk}
J.~G. Russo and A.~A. Tseytlin, ``{One loop four graviton amplitude in
  eleven-dimensional supergravity},'' {\em Nucl. Phys.} {\bf B508} (1997)
  245--259, \href{http://xxx.lanl.gov/abs/hep-th/9707134}{{\tt
  hep-th/9707134}}.

\bibitem{Chester:2018aa}
S.~M. Chester, S.~S. Pufu, and X.~Yin, ``{The M-Theory S-Matrix From ABJM:
  Beyond 11D Supergravity},'' \href{http://xxx.lanl.gov/abs/1804.00949}{{\tt
  1804.00949}}.

\bibitem{Penedones:2010ue}
J.~Penedones, ``{Writing CFT correlation functions as AdS scattering
  amplitudes},'' {\em JHEP} {\bf 03} (2011) 025,
  \href{http://xxx.lanl.gov/abs/1011.1485}{{\tt 1011.1485}}.

\bibitem{Kallosh:1998qs}
R.~Kallosh and A.~Rajaraman, ``{Vacua of M theory and string theory},'' {\em
  Phys. Rev.} {\bf D58} (1998) 125003,
  \href{http://xxx.lanl.gov/abs/hep-th/9805041}{{\tt hep-th/9805041}}.

\bibitem{Green:2008bf}
M.~B. Green, J.~G. Russo, and P.~Vanhove, ``{Modular properties of two-loop
  maximal supergravity and connections with string theory},'' {\em JHEP} {\bf
  07} (2008) 126, \href{http://xxx.lanl.gov/abs/0807.0389}{{\tt 0807.0389}}.

\bibitem{DHoker:2005vch}
E.~D'Hoker and D.~H. Phong, ``{Two-loop superstrings VI: Non-renormalization
  theorems and the 4-point function},'' {\em Nucl. Phys.} {\bf B715} (2005)
  3--90, \href{http://xxx.lanl.gov/abs/hep-th/0501197}{{\tt hep-th/0501197}}.

\bibitem{Gomez:2013sla}
H.~Gomez and C.~R. Mafra, ``{The closed-string 3-loop amplitude and
  S-duality},'' {\em JHEP} {\bf 10} (2013) 217,
  \href{http://xxx.lanl.gov/abs/1308.6567}{{\tt 1308.6567}}.

\bibitem{Alday:2014tsa}
L.~F. Alday, A.~Bissi, and T.~Lukowski, ``{Lessons from crossing symmetry at
  large N},'' {\em JHEP} {\bf 06} (2015) 074,
  \href{http://xxx.lanl.gov/abs/1410.4717}{{\tt 1410.4717}}.

\bibitem{Caron-Huot:2017vep}
S.~Caron-Huot, ``{Analyticity in Spin in Conformal Theories},'' {\em JHEP} {\bf
  09} (2017) 078, \href{http://xxx.lanl.gov/abs/1703.00278}{{\tt 1703.00278}}.

\bibitem{Osborn:1993cr}
H.~Osborn and A.~Petkou, ``{Implications of conformal invariance in field
  theories for general dimensions},'' {\em Annals Phys.} {\bf 231} (1994)
  311--362, \href{http://xxx.lanl.gov/abs/hep-th/9307010}{{\tt
  hep-th/9307010}}.

\bibitem{Bastianelli:2000hi}
F.~Bastianelli, S.~Frolov, and A.~A. Tseytlin, ``{Conformal anomaly of (2,0)
  tensor multiplet in six-dimensions and AdS / CFT correspondence},'' {\em
  JHEP} {\bf 02} (2000) 013, \href{http://xxx.lanl.gov/abs/hep-th/0001041}{{\tt
  hep-th/0001041}}.

\bibitem{Harvey:1998bx}
J.~A. Harvey, R.~Minasian, and G.~W. Moore, ``{NonAbelian tensor multiplet
  anomalies},'' {\em JHEP} {\bf 09} (1998) 004,
  \href{http://xxx.lanl.gov/abs/hep-th/9808060}{{\tt hep-th/9808060}}.

\bibitem{Intriligator:2000eq}
K.~A. Intriligator, ``{Anomaly matching and a Hopf-Wess-Zumino term in 6d,
  N=(2,0) field theories},'' {\em Nucl. Phys.} {\bf B581} (2000) 257--273,
  \href{http://xxx.lanl.gov/abs/hep-th/0001205}{{\tt hep-th/0001205}}.

\bibitem{Beccaria:2015ypa}
M.~Beccaria and A.~A. Tseytlin, ``{Conformal anomaly c-coefficients of
  superconformal 6d theories},'' {\em JHEP} {\bf 01} (2016) 001,
  \href{http://xxx.lanl.gov/abs/1510.02685}{{\tt 1510.02685}}.

\bibitem{Cordova:2015vwa}
C.~Cordova, T.~T. Dumitrescu, and X.~Yin, ``{Higher Derivative Terms, Toroidal
  Compactification, and Weyl Anomalies in Six-Dimensional (2,0) Theories},''
  \href{http://xxx.lanl.gov/abs/1505.03850}{{\tt 1505.03850}}.

\bibitem{Cordova:2015fha}
C.~Cordova, T.~T. Dumitrescu, and K.~Intriligator, ``{Anomalies,
  renormalization group flows, and the a-theorem in six-dimensional (1, 0)
  theories},'' {\em JHEP} {\bf 10} (2016) 080,
  \href{http://xxx.lanl.gov/abs/1506.03807}{{\tt 1506.03807}}.

\bibitem{Beccaria:2017dmw}
M.~Beccaria and A.~A. Tseytlin, ``{C$_{T}$ for higher derivative conformal
  fields and anomalies of (1, 0) superconformal 6d theories},'' {\em JHEP} {\bf
  06} (2017) 002, \href{http://xxx.lanl.gov/abs/1705.00305}{{\tt 1705.00305}}.

\bibitem{Yankielowicz:2017xkf}
S.~Yankielowicz and Y.~Zhou, ``{Supersymmetric R{\'e}nyi entropy and Anomalies
  in 6d (1,0) SCFTs},'' {\em JHEP} {\bf 04} (2017) 128,
  \href{http://xxx.lanl.gov/abs/1702.03518}{{\tt 1702.03518}}.

\bibitem{Chang:2017xmr}
C.-M. Chang and Y.-H. Lin, ``{Carving Out the End of the World or
  (Superconformal Bootstrap in Six Dimensions)},'' {\em JHEP} {\bf 08} (2017)
  128, \href{http://xxx.lanl.gov/abs/1705.05392}{{\tt 1705.05392}}.

\bibitem{Dolan:2004mu}
F.~A. Dolan, L.~Gallot, and E.~Sokatchev, ``{On four-point functions of 1/2-BPS
  operators in general dimensions},'' {\em JHEP} {\bf 0409} (2004) 056,
  \href{http://xxx.lanl.gov/abs/hep-th/0405180}{{\tt hep-th/0405180}}.

\bibitem{Heslop:2004du}
P.~J. Heslop, ``{Aspects of superconformal field theories in six dimensions},''
  {\em JHEP} {\bf 07} (2004) 056,
  \href{http://xxx.lanl.gov/abs/hep-th/0405245}{{\tt hep-th/0405245}}.

\bibitem{Ferrara:2001uj}
S.~Ferrara and E.~Sokatchev, ``{Universal properties of superconformal OPEs for
  1/2 BPS operators in $3 \leq d \leq 6$},'' {\em New J.Phys.} {\bf 4} (2002)
  2, \href{http://xxx.lanl.gov/abs/hep-th/0110174}{{\tt hep-th/0110174}}.

\bibitem{Nirschl:2004pa}
M.~Nirschl and H.~Osborn, ``{Superconformal Ward identities and their
  solution},'' {\em Nucl.Phys.} {\bf B711} (2005) 409--479,
  \href{http://xxx.lanl.gov/abs/hep-th/0407060}{{\tt hep-th/0407060}}.

\bibitem{Pope:1989ew}
C.~N. Pope, L.~J. Romans, and X.~Shen, ``{The Complete Structure of
  W(Infinity)},'' {\em Phys. Lett.} {\bf B236} (1990) 173--178.

\bibitem{Pope:1989sr}
C.~N. Pope, L.~J. Romans, and X.~Shen, ``{$W$(infinity) and the Racah-wigner
  Algebra},'' {\em Nucl. Phys.} {\bf B339} (1990) 191--221.

\bibitem{Campoleoni:2010zq}
A.~Campoleoni, S.~Fredenhagen, S.~Pfenninger, and S.~Theisen, ``{Asymptotic
  symmetries of three-dimensional gravity coupled to higher-spin fields},''
  {\em JHEP} {\bf 11} (2010) 007, \href{http://xxx.lanl.gov/abs/1008.4744}{{\tt
  1008.4744}}.

\bibitem{Bouwknegt:1988sv}
P.~Bouwknegt, ``{EXTENDED CONFORMAL ALGEBRAS},'' {\em Phys. Lett.} {\bf B207}
  (1988) 295. [,295(1988)].

\bibitem{Headrick:2015gba}
M.~Headrick, A.~Maloney, E.~Perlmutter, and I.~G. Zadeh, ``{R{\'e}nyi
  entropies, the analytic bootstrap, and 3D quantum gravity at higher genus},''
  {\em JHEP} {\bf 07} (2015) 059,
  \href{http://xxx.lanl.gov/abs/1503.07111}{{\tt 1503.07111}}.

\bibitem{Hornfeck:1992he}
K.~Hornfeck, ``{The Minimal supersymmetric extension of WA(n-1)},'' {\em Phys.
  Lett.} {\bf B275} (1992) 355--360.

\bibitem{Hornfeck:1993kp}
K.~Hornfeck, ``{Classification of structure constants for W algebras from
  highest weights},'' {\em Nucl. Phys.} {\bf B411} (1994) 307--320,
  \href{http://xxx.lanl.gov/abs/hep-th/9307170}{{\tt hep-th/9307170}}.

\bibitem{Blumenhagen:1994wg}
R.~Blumenhagen, W.~Eholzer, A.~Honecker, K.~Hornfeck, and R.~Hubel, ``{Coset
  realization of unifying W algebras},'' {\em Int. J. Mod. Phys.} {\bf A10}
  (1995) 2367--2430, \href{http://xxx.lanl.gov/abs/hep-th/9406203}{{\tt
  hep-th/9406203}}.

\bibitem{Linshaw:2017tvv}
A.~R. Linshaw, ``{Universal two-parameter $\mathcal{W}_{\infty}$-algebra and
  vertex algebras of type $\mathcal{W}(2,3,\dots, N)$},''
  \href{http://xxx.lanl.gov/abs/1710.02275}{{\tt 1710.02275}}.

\bibitem{Rastelli:2017ymc}
L.~Rastelli and X.~Zhou, ``{Holographic Four-Point Functions in the (2, 0)
  Theory},'' \href{http://xxx.lanl.gov/abs/1712.02788}{{\tt 1712.02788}}.

\bibitem{Aharony:2016dwx}
O.~Aharony, L.~F. Alday, A.~Bissi, and E.~Perlmutter, ``{Loops in AdS from
  Conformal Field Theory},'' {\em JHEP} {\bf 07} (2017) 036,
  \href{http://xxx.lanl.gov/abs/1612.03891}{{\tt 1612.03891}}.

\bibitem{Rastelli:2017udc}
L.~Rastelli and X.~Zhou, ``{How to Succeed at Holographic Correlators Without
  Really Trying},'' \href{http://xxx.lanl.gov/abs/1710.05923}{{\tt
  1710.05923}}.

\bibitem{Fitzpatrick:2011hu}
A.~L. Fitzpatrick and J.~Kaplan, ``{Analyticity and the Holographic
  S-Matrix},'' {\em JHEP} {\bf 10} (2012) 127,
  \href{http://xxx.lanl.gov/abs/1111.6972}{{\tt 1111.6972}}.

\bibitem{DHoker:2005jhf}
E.~D'Hoker, M.~Gutperle, and D.~H. Phong, ``{Two-loop superstrings and
  S-duality},'' {\em Nucl. Phys.} {\bf B722} (2005) 81--118,
  \href{http://xxx.lanl.gov/abs/hep-th/0503180}{{\tt hep-th/0503180}}.

\bibitem{Berkovits:2006vc}
N.~Berkovits, ``{New higher-derivative R**4 theorems},'' {\em Phys. Rev. Lett.}
  {\bf 98} (2007) 211601, \href{http://xxx.lanl.gov/abs/hep-th/0609006}{{\tt
  hep-th/0609006}}.

\bibitem{Heemskerk:2009pn}
I.~Heemskerk, J.~Penedones, J.~Polchinski, and J.~Sully, ``{Holography from
  Conformal Field Theory},'' {\em JHEP} {\bf 10} (2009) 079,
  \href{http://xxx.lanl.gov/abs/0907.0151}{{\tt 0907.0151}}.

\bibitem{Heslop:2017sco}
P.~Heslop and A.~E. Lipstein, ``{M-theory Beyond The Supergravity
  Approximation},'' {\em JHEP} {\bf 02} (2018) 004,
  \href{http://xxx.lanl.gov/abs/1712.08570}{{\tt 1712.08570}}.

\bibitem{Alday:2017vkk}
L.~F. Alday and S.~Caron-Huot, ``{Gravitational S-matrix from CFT dispersion
  relations},'' \href{http://xxx.lanl.gov/abs/1711.02031}{{\tt 1711.02031}}.

\bibitem{Chester:2018lbz}
S.~M. Chester, ``{AdS$_4$/CFT$_3$ for Unprotected Operators},''
  \href{http://xxx.lanl.gov/abs/1803.01379}{{\tt 1803.01379}}.

\bibitem{Arutyunov:2002ff}
G.~Arutyunov and E.~Sokatchev, ``{Implications of superconformal symmetry for
  interacting (2,0) tensor multiplets},'' {\em Nucl. Phys.} {\bf B635} (2002)
  3--32, \href{http://xxx.lanl.gov/abs/hep-th/0201145}{{\tt hep-th/0201145}}.

\bibitem{ElShowk:2012hu}
S.~El-Showk and M.~F. Paulos, ``{Bootstrapping Conformal Field Theories with
  the Extremal Functional Method},'' {\em Phys. Rev. Lett.} {\bf 111} (2013),
  no.~24 241601, \href{http://xxx.lanl.gov/abs/1211.2810}{{\tt 1211.2810}}.

\bibitem{Agmon:2017xes}
N.~B. Agmon, S.~M. Chester, and S.~S. Pufu, ``{Solving M-theory with the
  Conformal Bootstrap},'' \href{http://xxx.lanl.gov/abs/1711.07343}{{\tt
  1711.07343}}.

\bibitem{Green:2005ba}
M.~B. Green and P.~Vanhove, ``{Duality and higher derivative terms in M
  theory},'' {\em JHEP} {\bf 01} (2006) 093,
  \href{http://xxx.lanl.gov/abs/hep-th/0510027}{{\tt hep-th/0510027}}.

\bibitem{Green:2006gt}
M.~B. Green, J.~G. Russo, and P.~Vanhove, ``{Non-renormalisation conditions in
  type II string theory and maximal supergravity},'' {\em JHEP} {\bf 02} (2007)
  099, \href{http://xxx.lanl.gov/abs/hep-th/0610299}{{\tt hep-th/0610299}}.

\bibitem{Bjornsson:2010wm}
J.~Bjornsson and M.~B. Green, ``{5 loops in 24/5 dimensions},'' {\em JHEP} {\bf
  08} (2010) 132, \href{http://xxx.lanl.gov/abs/1004.2692}{{\tt 1004.2692}}.

\bibitem{Vanhove:2010nf}
P.~Vanhove, ``{The Critical ultraviolet behaviour of N=8 supergravity
  amplitudes},'' \href{http://xxx.lanl.gov/abs/1004.1392}{{\tt 1004.1392}}.

\bibitem{Wang:2015aua}
Y.~Wang and X.~Yin, ``{Supervertices and Non-renormalization Conditions in
  Maximal Supergravity Theories},''
  \href{http://xxx.lanl.gov/abs/1505.05861}{{\tt 1505.05861}}.

\bibitem{Bern:2018jmv}
Z.~Bern, J.~J. Carrasco, W.-M. Chen, A.~Edison, H.~Johansson,
  J.~Parra-Martinez, R.~Roiban, and M.~Zeng, ``{Ultraviolet Properties of N = 8
  Supergravity at Five Loops},'' \href{http://xxx.lanl.gov/abs/1804.09311}{{\tt
  1804.09311}}.

\bibitem{Buican:2016hpb}
M.~Buican, J.~Hayling, and C.~Papageorgakis, ``{Aspects of Superconformal
  Multiplets in D>4},'' {\em JHEP} {\bf 11} (2016) 091,
  \href{http://xxx.lanl.gov/abs/1606.00810}{{\tt 1606.00810}}.

\bibitem{Gopakumar:2016cpb}
R.~Gopakumar, A.~Kaviraj, K.~Sen, and A.~Sinha, ``{A Mellin space approach to
  the conformal bootstrap},'' {\em JHEP} {\bf 05} (2017) 027,
  \href{http://xxx.lanl.gov/abs/1611.08407}{{\tt 1611.08407}}.

\bibitem{Green:1987sp}
M.~B. Green, J.~H. Schwarz, and E.~Witten, {\em {SUPERSTRING THEORY. VOL. 1:
  INTRODUCTION}}.
\newblock Cambridge Monographs on Mathematical Physics. 1988.

\bibitem{Green:1997tv}
M.~B. Green and M.~Gutperle, ``{Effects of D instantons},'' {\em Nucl. Phys.}
  {\bf B498} (1997) 195--227,
  \href{http://xxx.lanl.gov/abs/hep-th/9701093}{{\tt hep-th/9701093}}.

\end{thebibliography}\endgroup

\end{document}